\documentclass[12pt,preprint]{aastex}

\begin{document}

\title{Examining the Effect of the Map-Making Algorithm on Observed Power Asymmetry in WMAP Data}

\author{P.~E.~Freeman\altaffilmark{1}, C.~R.~Genovese\altaffilmark{1}, C.~J.~Miller\altaffilmark{2}, R.~C.~Nichol\altaffilmark{3}, \& L.~Wasserman\altaffilmark{1}}
\altaffiltext{1}{Department of Statistics, Carnegie Mellon University, 5000 Forbes Ave., Pittsburgh, PA~~15213}
\altaffiltext{2}{CTIO/NOAO, 950 North Cherry Avenue, Tucson, AZ~~85719}
\altaffiltext{3}{Institute of Cosmology and Gravitation, University of Portsmouth, Portsmouth, PO1 2EG, UK}
\email{pfreeman@cmu.edu}

\begin{abstract}
We analyze first-year data of {\it WMAP}
to determine the significance of asymmetry in summed power between
arbitrarily defined opposite hemispheres.  
We perform this analysis on maps that we create ourselves from
the time-ordered data, using software developed
independently of the {\it WMAP} team.
We find that over the multipole range
$l$~=~[2,64], the significance of asymmetry is $\sim$~10$^{-4}$, a
value insensitive to both frequency and power spectrum.
We determine the smallest multipole ranges exhibiting
significant asymmetry, and find twelve, including $l$~=~[2,3] and [6,7],
for which the significance $\rightarrow$ 0.
Examination of the twelve ranges indicates both an improbable
association between the direction of maximum significance and the ecliptic
plane (significance $\sim$~0.01), and that contours of least 
significance follow great circles inclined relative to the ecliptic 
at the largest scales.  The great circle for $l$~=~[2,3] passes over previously
reported preferred axes and is insensitive to frequency,
while the great circle for $l$~=~[6,7] is aligned
with the ecliptic poles.  
We examine how changing map-making parameters,
e.g., foreground masking, affects asymmetry.
Only one change appreciably reduces asymmetry:
asymmetry at large scales ($l \leq$~7) is rendered insignificant
if the magnitude of the {\it WMAP} dipole vector
(368.11 km s$^{-1}$) is increased 
by $\approx$ 1-3$\sigma$ ($\approx$ 2-6 km s$^{-1}$).
While confirmation of this result requires the recalibration of the 
time-ordered data, such a systematic change would be 
consistent with observations of frequency-independent asymmetry.
We conclude that the use of an incorrect dipole vector, in
combination with
a systematic or foreground process associated with the ecliptic,
may help to explain the observed power asymmetry.
\end{abstract}


\section{Introduction}

\setcounter{footnote}{3}

One of the most intriguing results gleaned from the 
intense study of the first-year data of the
{\it Wilkinson Microwave Anisotropy Probe} ({\it WMAP})\footnote{\scriptsize
\tt http://lambda.gsfc.nasa.gov/product/map/m\_products.cfm}
is the possible detection of
irregularities in the temperature field of
the Cosmic Microwave Background (CMB).
One interpretation is that the temperature field
is not a Gaussian random field, i.e.,
when it is expanded in terms of spherical harmonics
\begin{eqnarray}
T(\theta,\phi)~=~\sum_{l=0}^{\infty} \sum_{m=-l}^{l} a_{lm} Y_{lm}(\theta,\phi) \,, \nonumber
\end{eqnarray}
the modes $a_{lm}$ are not Gaussian random variables with zero mean and 
variance $<C_l^2>$, where
\begin{eqnarray}
C_l~=~\frac{1}{2l+1}\sum_{m=-l}^{l} {\vert}a_{lm}{\vert}^2 \,. \nonumber
\end{eqnarray}
Standard inflationary cosmology predicts a random Gaussian
field.  Thus non-Gaussianity, if it exists,\footnote{
See Magueijo \& Medeiros and references therein for the
instructive example of non-Gaussianity detected
in {\it Cosmic Background Explorer} data.} 
leads to the conclusion that
the standard inflationary paradigm is incomplete or
incorrect (see, e.g., {\S}2 of Copi, Huterer, \& Starkman 2004 and references
therein).  However, there are other plausible interpretations for 
temperature field irregularities: they may indicate unknown sources of 
foreground emission in the CMB bandpass; they may be
an artifact of an unknown problem with the {\it WMAP} instrument; or 
they may be an artifact 
of the algorithm by which the temperature field maps are generated.

A number of different approaches have been applied in the search for
irregularities in the {\it WMAP} data.
Komatsu et al.~(2003) use techniques based on both the angular bispectrum
and Minkowski functionals to determine that the {\it WMAP} data are
consistent with Gaussianity, while Patanchon et al.~(2004) apply a
blind multi-component analysis and achieve the same conclusion, while
at the same time finding evidence of weak residual foreground emission 
in the {\it WMAP} Q-band.  However, Coles et al.~(2004) examine phase
correlations and conclude that there are departures from uniformity
probably caused by the foreground, while Vielva et al.~(2004) and
McEwen et al.~(2004) use wavelet-based techniques and examine the
skewness and kurtosis of wavelet coefficients, finding deviations
from Gaussianity of significance\footnote{
Throughout this paper, significance is given as the tail
integral $\alpha$~=~$\int_{S_o}^{\infty} p_{\rm null}$, where
$p_{\rm null}$ is a sampling distribution for statistic $S$ given
that the null hypothesis is true, and $S_o$ is the observed
statistic.  Conventionally, the alternate hypothesis is accepted
if $\alpha \leq$~0.05, although the standard of acceptance is a
subjective choice.  Note that the phrases ``more significant" and
``less significant" refer to the value $\alpha$ being {\it smaller}
and {\it larger} respectively.
}
$\alpha$~=~0.047 and 0.017, respectively.
(The former number is McEwen et al.'s revised rendering of Vielva et al.'s
result; Vielva et al.~claim $\alpha$~=~0.001.)  Vielva et al.
examine the frequency dependence of their non-Gaussian signal and
conclude that systematic and foreground effects can be ruled out.
Copi et al.~and Land \& Magueijo (2005a) use 
multipole vectors to examine the lowest 
multipoles ($l$~=~[2,8]) and find correlations that are unlikely in Gaussian
random fields at the $\alpha \sim$ 0.01 and 0.001 level respectively.
De Oliveira-Costa et al.~(2004) and Schwarz et al.~(2004) examine
the unusual alignments of the quadrupole and octopole planes; the
latter concludes that the alignments are unlikely at the
$\alpha$~=~10$^{-3}$ level, and note that three of the four planes are
orthogonal to the ecliptic plane, while the fourth is orthogonal
to the supergalactic plane, 
supporting the hypothesis that unmodeled foregrounds are a tangible
cause of irregularities in the data.
Vale (2005) proposes that these improbable alignments are a by-product
of another foreground contamination mechanism,
the weak lensing of the CMB dipole by local large-scale structures.

Other researchers have looked at differences between 
standard hemispheres (galactic and ecliptic)
and arbitrarily defined hemispheres.
Eriksen et al.~(2004a; hereafter Eriksen I) and 
Hansen, Banday, \& G\'orski (2004a; hereafter Hansen I) use
co-added V- and W-band {\it WMAP} radiometer maps
to find asymmetry in summed power, i.e., the ratio of 
summed powers is significantly different from one
($\alpha \sim$ 10$^{-3}$).
They find that the north pole (NP) direction for maximum
asymmetry shifts from the galactic NP toward the galactic
plane as the lowest $l$ values are excluded from the analysis,
and they conclude that the orientation of the NP at higher $l$
values indicates that residual foreground contamination is
unlikely.  
Eriksen et al.~(2004b) follow up this work using
Minkowski functionals and skeleton length to ascertain 
non-Gaussianity in the northern Galactic hemisphere at 
the $\alpha \sim$ 10$^{-3}$ level.  Park (2004) uses
a similar mathematical framework and finds a difference
between the northern and southern Galactic hemispheres
and an asymmetry in the southern hemisphere, both
significant at the $\alpha \approx$ 0.01 level.
Hansen et al.~(2004b) use a local-curvature method and find
non-Gaussian features when considering the northern and southern
Galactic hemispheres separately at the $\alpha \sim$ 0.01 level, and
find that maximum asymmetry occurs when the NP is close
to north ecliptic pole, suggesting a foreground effect.
Larson \& Wandelt (2004) examine extrema outside the most 
conservative {\it WMAP} foreground mask (Kp0 mask) and 
in addition to rejecting the Gaussian hypothesis on the whole
sky, they find the variance of maxima and minima to be low 
($\alpha$~=~0.01) in the Ecliptic northern hemisphere but consistent
with the null in the southern hemisphere.

These results indicate that the evidence for irregularities in
the {\it WMAP} data is
tantalizing, though ambiguous, and that there is as yet
no consensus as to its root cause.  However, the hints of 
special alignments of hemispheres for which power asymmetry
is maximized, as well as special alignments of the quadrupole
and octopole planes, suggest that an incomplete knowledge
of foregrounds may play a leading role.
(We note the effect of one possible foreground 
contamination mechanism, the Sunyaev-Zel'dovich effect, was recently
discounted by Hansen et al.~2005, in accordance with Bennett et al.~2003b.)

One heretofore unexamined aspect of this problem
is the role of the map-making algorithm itself.  The authors
listed above have worked with the first-year data that is provided
by the {\it WMAP} team, and thus have not examined how the choices
made by the {\it WMAP} team in making maps 
(Hinshaw et al.~2003a, hereafter Hinshaw I) affect the evidence
for irregularities in the data.  
For this reason,
we have independently developed map-making software that operates
on the first-year calibrated time-ordered {\it WMAP} data, and
we analyze the resulting maps to determine: (a) if there is statistically 
significant asymmetry between arbitrarily defined hemispheres;
and (b) if altering the map-making algorithm affects
the observed results. 

In {\S}\ref{sect:map}, we review the basics of map-making and discuss our own 
map-making algorithm, which we have created through the study of 
Hinshaw I and discussions with the {\it WMAP} team.  In {\S}\ref{sect:powspec},
we analyze foreground-corrected, co-added Q-, V-, and/or W-band maps 
to determine
asymmetry as a function of direction and to estimate significance
both as a function of direction and globally over the whole sphere.
In {\S}\ref{sect:mapassume}, we determine how the distribution of
observed asymmetry values and the direction and significance of the 
maximum value are affected by altering the map-making algorithm.
In {\S}\ref{sect:conc}, we provide a summary and conclusions, while
in the Appendix we provide complete details of our map-making recipe.

\section{Map-Making Algorithm}

\label{sect:map}

\subsection{Paradigm}

\label{sect:map_prelim}

We begin by reviewing the theory underlying the making of maps for
{\it WMAP}; we direct the interested
reader to Wright, Hinshaw, \& Bennett (1996), Wright (1996), 
Tegmark (1997), and Hinshaw I for more detail.
The goal of map-making is to determine the maximum likelihood
estimate of the true sky map ${\vec T}_{\rm p}$
(where the subscript $p$ indicates data as a function of sky pixel),
which is associated with the time-ordered data 
${\vec T}_t$ via the relation
\begin{eqnarray}
{\vec T}_t~=~{\bf M}{\vec T}_p + {\vec n}_t \,,
\nonumber
\end{eqnarray}
where the subscript $t$ indicates data as a function of time,
${\bf M}$ is the pixel-to-time mapping matrix, and
${\vec n}_t$ is the vector of samples from the noise distribution.

To simplify the problem, it is assumed that $n_t \sim N(0,\sigma_{\rm o}^2)$, 
i.e., each noise sample is independent and the sampling distribution is
the normal distribution with time-independent variance $\sigma_{\rm o}^2$.
(This is not actually true of the noise in raw {\it WMAP} data, and
ridding the data of the effects of 1/$f$ noise is a major component of the
data calibration process; see {\S}2.3.2 of Hinshaw I.)
With this assumption, the maximum-likelihood estimate is:
\begin{eqnarray}
{\vec T}_p~=~({\bf M}^{\rm T} {\bf C}^{-1} {\bf M})^{-1} ({\bf M}^{\rm T} {\bf C}^{-1} {\vec T}_t) \,. \nonumber
\end{eqnarray}
One may follow Tegmark (1997) and approximate the noise
covariance matrix ${\bf C}^{-1}$ with the
identity matrix ${\bf I}$, and one may furthermore assume that the 
matrix product
${\bf M}^{\rm T} {\bf M}$ is diagonally dominant, so that one does not have to
invert a $N_p \times N_p$ matrix (where $N_p$ can be as large as 
$\approx$ 3 $\times$ 10$^7$).
(Thus in first-year data processing, the {\it WMAP} team assumes the
beam response for each radiometer is a $\delta$-function.)
For the specific case of {\it WMAP}, the off-diagonal elements have 
magnitude $\approx$ 0.3\% relative to
the diagonal elements; this is the inverse of the total number of pixels that
may be paired with a given pixel, given the beam separation of
{\it WMAP} radiometers.
With this assumption in place,
\begin{eqnarray}
({\bf M}^{\rm T}{\bf M})^{-1} \approx {\vec n}_{\rm obs}^{-1} \,, \nonumber
\end{eqnarray}
where $n_{{\rm obs},p}$ is the number of times that sky pixel $p$ is
observed.

\subsection{Algorithm}

\label{sect:map_algor}

{\it WMAP} is comprised of ten differential radiometers
covering five frequency bands:  
K1 (20-25 GHz); Ka1 (28-36 GHz); Q1, Q2 (35-46 GHz);
V1, V2 (53-69 GHz); and W1, W2, W3, W4 (82-106 GHz).  The lowest frequency
radiometers are meant to assist the modelling of Galactic foreground
emission, while the highest frequency radiometers are the more important
ones for modelling the CMB.  (In this work, we concentrate upon
the data of the Q, V, and W bands.)
Each radiometer consists of two horns (denoted $A$ and $B$ by
Hinshaw I) that are
separated by $\approx$ 140$^{\circ}$.
Temperature differences ${\Delta}T_{{\rm raw},t}$
between two points on the sky are measured every 1.536/$n$ s,
where $n$ equals 12, 15, 20, and 30 for the K and Ka, Q, V, and W
bands respectively.\footnote{
There are four data associated with each temperature
difference:
\begin{eqnarray}
{\Delta}T_{{\rm raw},t}~=~T_{A,t}-T_{B,t}~=~\frac{1}{2}(d_{A3}+d_{A4})+\frac{1}{2}(d_{B3}+d_{B4}) \,, \nonumber
\end{eqnarray} 
where 3 and 4 denote linear orthogonal polarization modes.
These data are not available in the first-year release.
(To match the notation of equation 1 of Hinshaw I, replace A and B
with 1 and 2, respectively.)}
(Here, we change notation from $T$ to ${\Delta}T$, because our
interest lies in anisotropies.)
As described in Hinshaw I, raw temperature differences
are calibrated, to correct for varying baselines and gains
(and to correct for 1/$f$ noise).  As the first-year data release
contains only the calibrated differences ${\Delta}T_{{\rm cal},t}$,
these are what we analyze.

Here we summarize our map-making algorithm; more detail may be found
in the Appendix.

We begin with a zero-temperature map
and estimate ${\Delta}T_{{\rm CMB},p,{\rm true}}$ via an iterative
process.  During each iteration, the time-ordered data 
${\Delta}T_{{\rm cal},t}$ are treated 
sequentially.
\begin{itemize}
\item For each datum, we estimate the radiometer normal vectors
${\vec n}_A$ and ${\vec n}_B$.  The spacecraft orientation as a function
of time is encoded using quaternions
($q_1+q_2{\hat i}+q_3{\hat j}+q_4{\hat k}$ in scalar-vector form) that
are recorded at the beginning of each
1.536 s science frame.  Interpolation of quaternion elements
is crucial since the
spacecraft spins once every 129 s, or $\approx$ 4$^{\circ}$
per frame.
The {\it WMAP} team uses a Lagrange interpolating cubic polynomial that
is anchored to the
four points used in its determination.  The interpolated elements are
used to determine ${\vec n}_A$ and ${\vec n}_B$ in
galactic coordinates, and also to determine the
two sky pixels $p_A$ and $p_B$ that are associated with the datum.
\item The contribution of the Doppler-shifted monopole is removed from each
datum:
\begin{eqnarray} 
{\Delta}T_{{\rm CMB},t}~=~{\Delta}T_{{\rm cal},t}-T_o \times \left[ {\vec \beta}\cdot({\vec n}_A-{\vec n}_B) + ({\vec \beta}\cdot{\vec n}_A)^2 - ({\vec \beta}\cdot{\vec n}_B)^2 \right] \,,
\label{eqn:trans}
\end{eqnarray}
where $T_o$~=~2.725 K (Mather et al.~1999),
and ${\vec \beta}$ is the velocity of the 
satellite with
respect to the CMB rest frame expressed as a fraction of $c$.\footnote{
The effect of the second-order term, which the {\it WMAP} team
does not include in their map calculation, is $\lesssim$ 2$\mu$K.
See Figure \ref{fig:diffmapsnodipole}.
}
We linearly interpolate the barycentric velocity of {\it WMAP},
which is recorded every 30 science frames (46.08 s).
We assume the velocity of the Sun relative to the CMB rest frame 
in galactic coordinates to be 
$(v_{\odot,\rm gal}^x,v_{\odot,\rm gal}^y,v_{\odot,\rm gal}^z)$~=~($-$26.26,$-$243.71,274.63) km s$^{-1}$ (or $v_{\odot,\rm gal}$~=~368.11 
km s$^{-1}$; based on {\S}7.1 of Bennett et al.~2003a, hereafter Bennett I).
\item Running sums are continuously updated to estimate
${\Delta}T_{{\rm CMB},p}$ during the $(i+1)^{\rm th}$ iteration:
\begin{eqnarray}
n_{{\rm obs},p_A} &=& n_{{\rm obs},p_A}+w_t \label{eqn:na} \\
{\Delta}T_{{\rm CMB},p_A,i+1} &=& {\Delta}T_{{\rm CMB},p_A,i+1} + w_t\frac{{\Delta}T_{{\rm CMB},t}+(1-x_{\rm im}){\Delta}T_{{\rm CMB},p_B,i}}{1+x_{\rm im}} \label{eqn:ta}
\end{eqnarray}
and
\begin{eqnarray}
n_{{\rm obs},p_B} &=& n_{{\rm obs},p_B}+w_t \label{eqn:nb} \\
{\Delta}T_{{\rm CMB},p_B,i+1} &=& {\Delta}T_{{\rm CMB},p_B,i+1} + w_t\frac{-{\Delta}T_{{\rm CMB},t}+(1+x_{\rm im}){\Delta}T_{{\rm CMB},p_A,i}}{1-x_{\rm im}} \,, \label{eqn:tb}
\end{eqnarray}
where $x_{\rm im}$ is a loss-imbalance parameter tabulated
for each radiometer horn ($\sim$ 10$^{-3}$-10$^{-2}$;
Table 3 of Jarosik et al.~2003),
and $w_t$ is a statistical weight based on instrument temperature
that is expected
to differ from 1.0 by only $\pm$ 1\% (G.~Hinshaw 2004, private communication).
We assume $w_t$~=~1.

There are three situations in which the running sums are
not updated.  
First, if the data are flagged as bad, or
second, if there is a planet within $\theta_{\rm cut}$ degrees of 
either radiometer A or B, then {\it neither} running sum is updated.
(The {\it WMAP} team assumes $\theta_{\rm cut}$~=~1.5$^{\circ}$.)  
Third, if either 
radiometer is pointing to within the Galactic mask
the running sum for the opposite
radiometer is not updated.  This is to mitigate the effect of 
elliptical radiometer beams, which cause the differential signal
associated with bright sources to change with spacecraft orientation
and thus to bias map estimates (Hinshaw I, {\S}3.3.4).
\end{itemize}

After all the time-ordered data are treated, the weighted average
of ${\Delta}T_{{\rm CMB},p,i+1}$ is determined by dividing by
$n_{{\rm obs},p}$; this estimate is passed to the next iteration.

The iterative process determines the spatial distribution of relative
temperature differences, and does not determine a zero point.  As noted
in the Appendix, the {\it WMAP} team determines a zero point via model
fitting.  We assume the {\it WMAP} zero-point values, i.e., we apply a constant
offset to each pixel such that our mean temperature matches that determined
by {\it WMAP}.

\subsection{Example Maps}

\label{sect:map_examples}

\begin{figure}[p]
\centering
\includegraphics[width=4.0in]{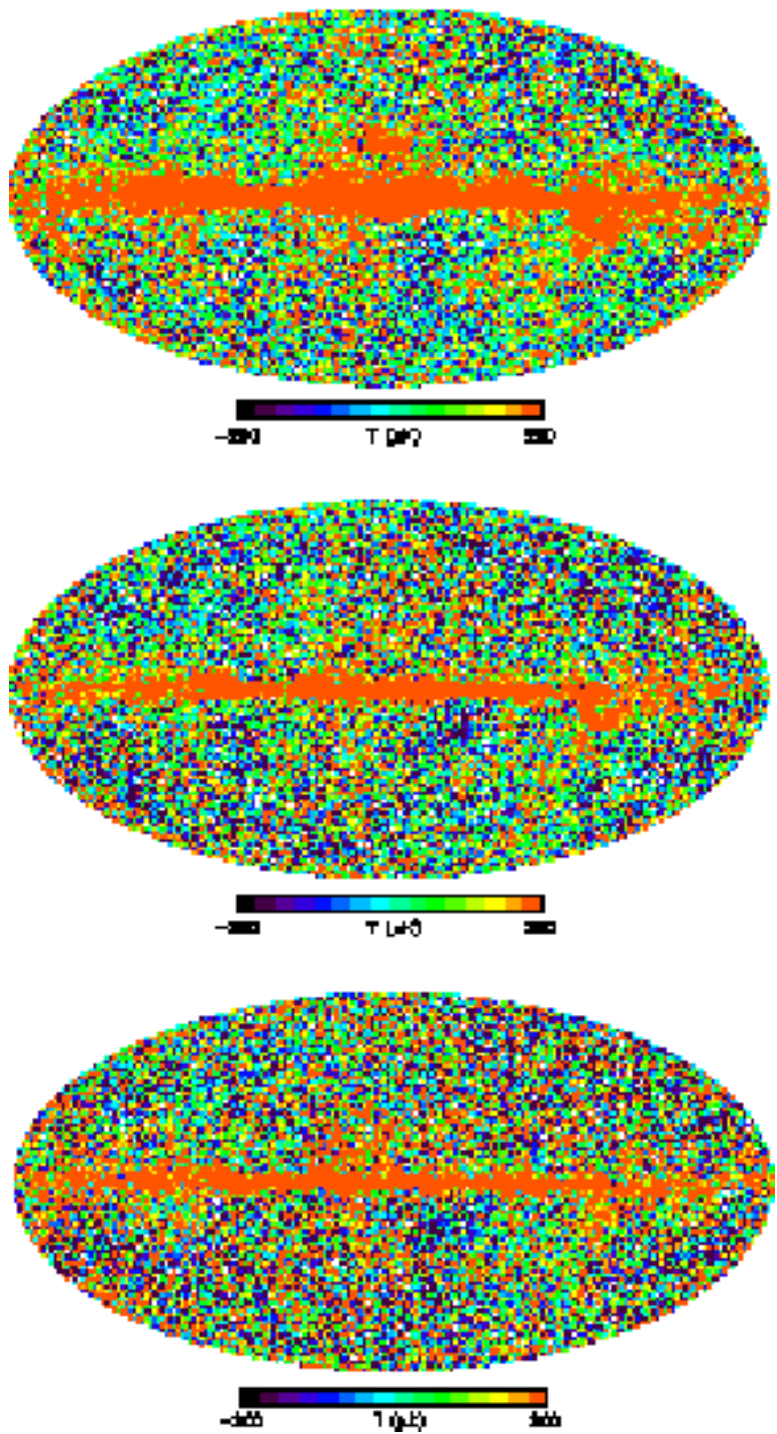}
\caption{From top to bottom, combined Q-, V-, and W-band radiometer maps
created by making maps for each individual radiometer and summing them
using the weighting scheme of equation (24) of Hinshaw II.}
\label{fig:rawmaps}
\end{figure}

We display combined Q-, V-, and W-band radiometer maps in 
Figure \ref{fig:rawmaps}.  Maps are first created for each individual
radiometer, then combined using the weighting scheme of
equation (24) of Hinshaw et al.~(2003b; hereafter Hinshaw II).  

\begin{figure}[p]
\centering
\includegraphics[width=4.0in]{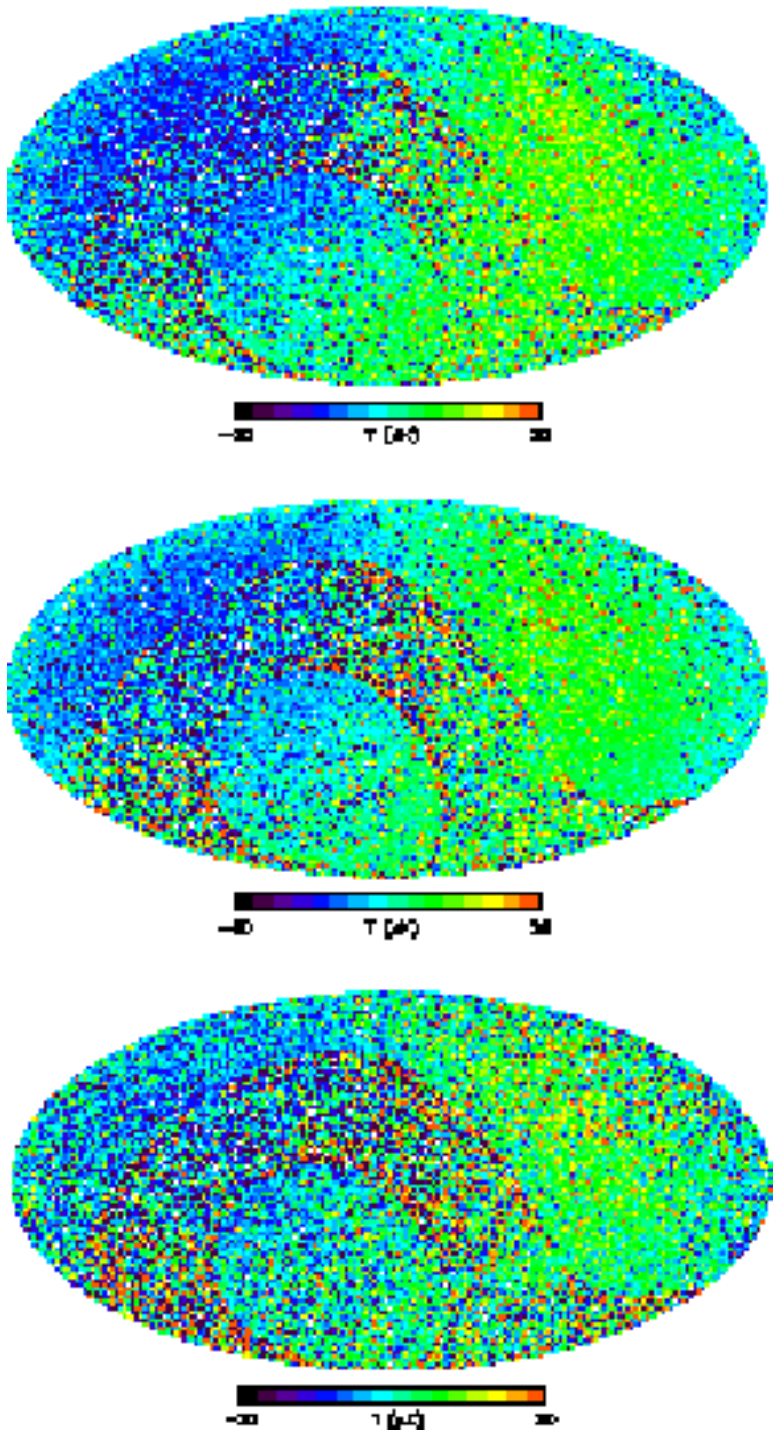}
\caption{From top to bottom, combined Q-, V-, and W-band radiometer difference 
maps (${\Delta}T_{\rm WMAP} - {\Delta}T$) created with 
combined {\it WMAP} maps and 
the maps shown in Figure \ref{fig:rawmaps}.
There are two
structures present in these data: a speckled pattern related to 
the statistical weighting of first day data; and a dipolar 
structure of amplitude $\approx$ 7 $\mu$K in the direction
$(l_{\rm gal},b_{\rm gal}) \approx$ ($-$86$^{\circ}$,$-$19$^{\circ}$).
See also Figure \ref{fig:diffmapshist}.}
\label{fig:diffmaps}
\end{figure}

While our maps compare very favorably
with their counterparts shown in Figure 2 of Bennett I, they do not
match exactly.
In Figure \ref{fig:diffmaps}, we show the differences between combined
Q-, V-, and W-band {\it WMAP} maps and our combined maps.  Two residual
structures are apparent in this figure.
The first is a speckled pattern aligned with the scan pattern
for the first day of observations.  Its cause is unknown: while it may be
a thermal effect manifested as significant departures
($\sim$ 10\%) of $w_t$ from unity, a timing mismatch between datasets
has also been suggested (Hinshaw, private communication).
The fact that the values of $w_t$ are not
included in the data release makes it more difficult to determine the cause;
we would ask that these values be included in future data releases.
In this particular instance, the effect of the missing data
is not grievous; we have examined difference maps made
with and without the first day's data and determined that the
speckled pattern has negligible effect on maps and asymmetry values.
Since our bias is to avoid arbitrary
data cuts when possible, we include the first day's data in
all analyses discussed below.

The second residual structure is a dipole that has amplitude
$\approx$ 7$\mu$K and direction $(l_{\rm gal},b_{\rm gal}) \approx$
($-$86$^{\circ}$,$-$19$^{\circ}$) in all three bands, as determined by
the {\tt HEALPix}\footnote{
The {\tt HEALPix} software is described in
G\'orski, Hivon, \& Wandelt (1999) and is available at\\
{\scriptsize \tt http://www.eso.org/science/healpix}.}
IDL routine {\tt remove\_dipole}.\footnote{
We specify galactic longitude $l$ and latitude $b$ with a subscript
``gal" so as not to confuse longitude with the multipole number $l$.}
The appearance of residual dipole structures is not surprising.
When calibrating the data,
the {\it WMAP} team assumed the barycentric velocity vector estimated from
{\it COBE} DMR data,
$(v_{\odot,\rm gal}^x,v_{\odot,\rm gal}^y,v_{\odot,\rm gal}^z)$~=~($-$24.57,$-$244.47,275.00) km s$^{-1}$
(G.~Hinshaw, private communication; based on Bennett et al.~1996, in
which the vector elements are not explicitly stated), and in Bennett I
it is mentioned that residual dipole structures are observed and are used
to improve estimates of the dipole temperature and direction.  
We conclude that the residual dipole that we observe
reflects a combination of
numerical differences the dipole unit vectors and statistical weights
that we use and those assumed by the {\it WMAP} team.
For completeness, we determine that removing the speckled pattern 
by not using the
first day's data has negligible effect on the residual dipole amplitude
($\lesssim$ 0.5\%).

\begin{figure}[p]
\centering
\includegraphics[width=4.0in]{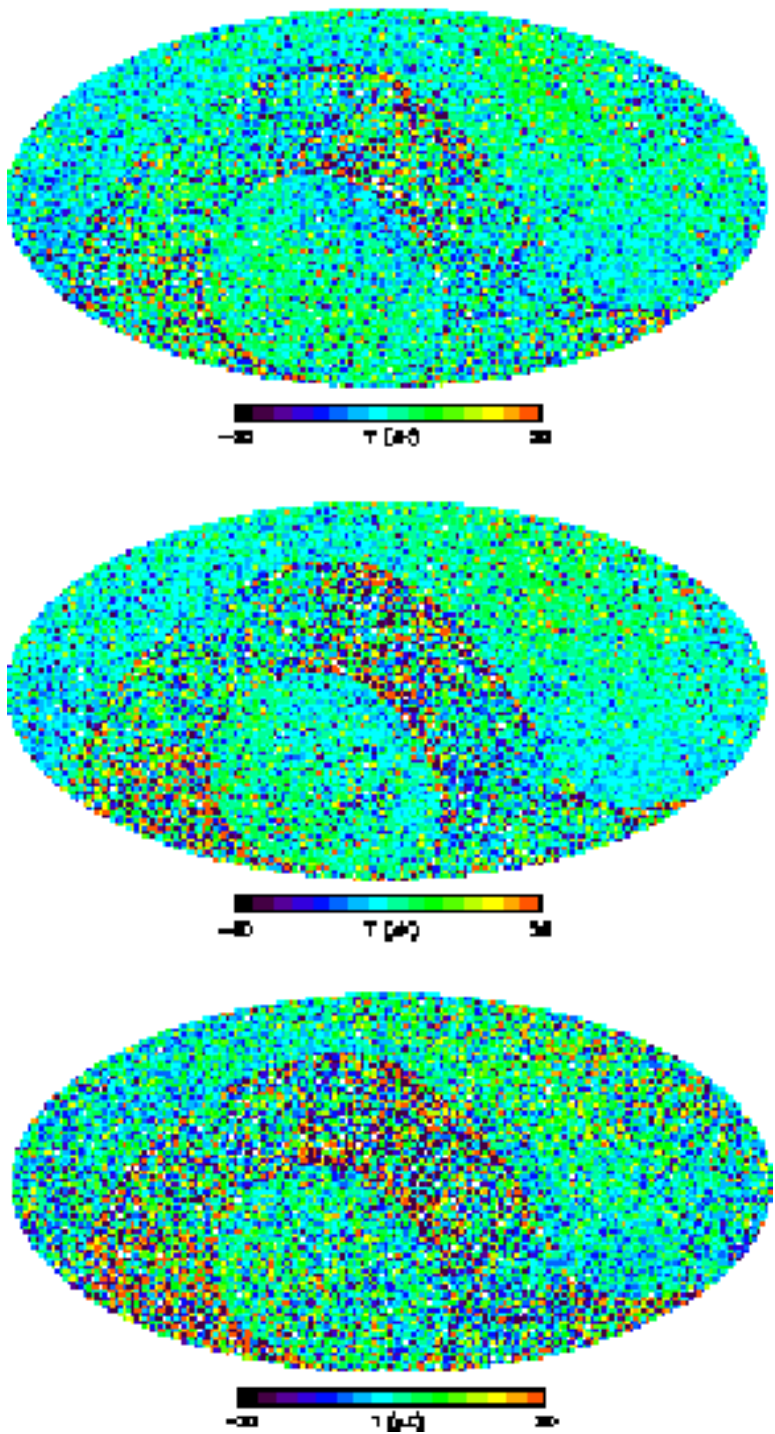}
\caption{Same as Figure \ref{fig:diffmaps}, but with the residual dipolar
structure removed.  The remaining structure is due to our including
the second-order term when computing the Doppler shift of the monopole; 
the {\it WMAP} team does not include this term in their calculations.
See also Figure \ref{fig:diffmapshist}.}
\label{fig:diffmapsnodipole}
\end{figure}

In Figure \ref{fig:diffmapsnodipole}, we show the difference maps of 
Figure \ref{fig:diffmaps} with the residual dipoles removed.  
Another
structure, aligned with the CMB dipole, is apparent in these figures;
this is the effect of including the second-order term when computing
the Doppler shift of the monopole, which the {\it WMAP} team
does not include in their calculations.

\begin{figure}[p]
\centering
\includegraphics[width=3.0in]{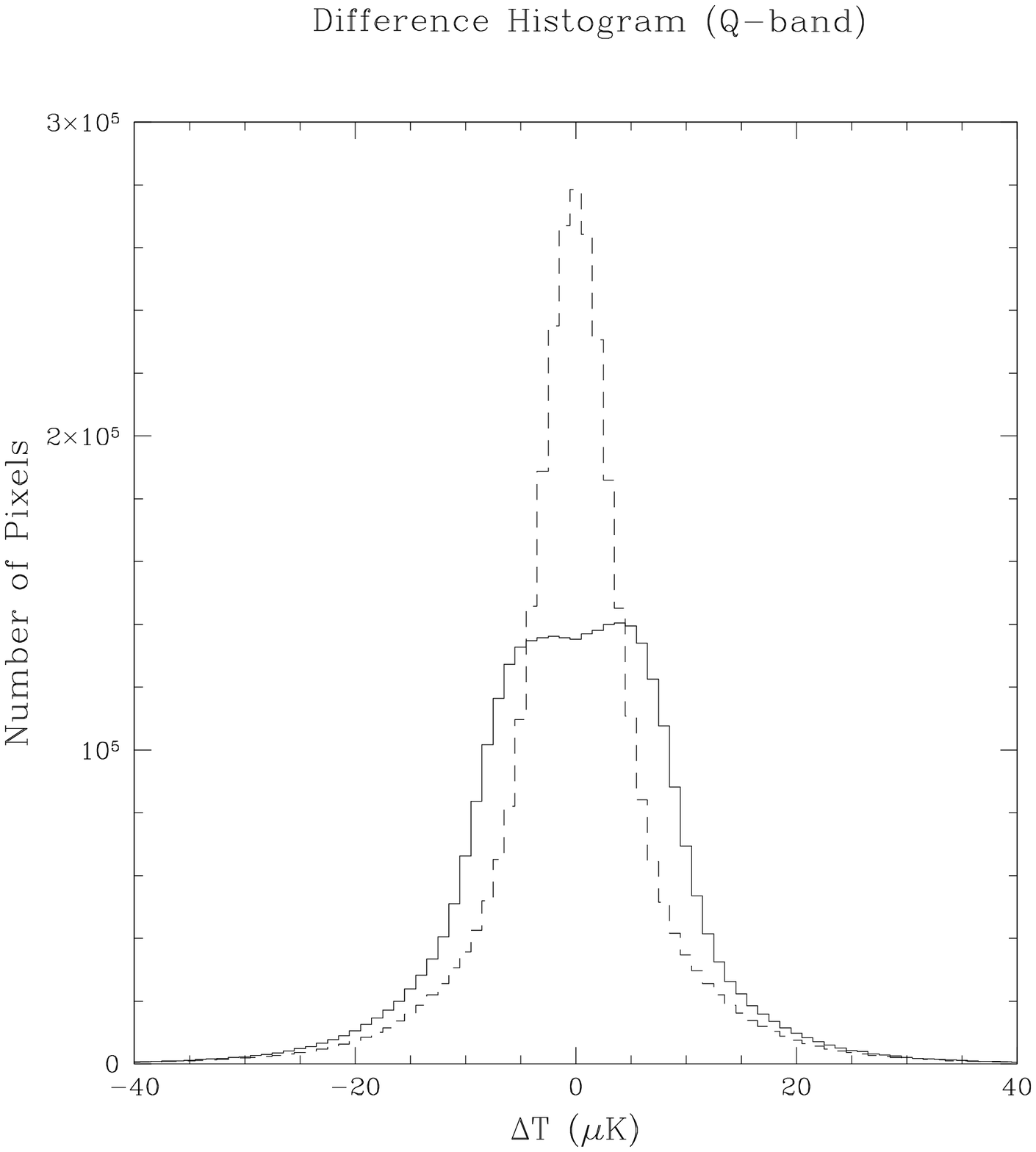}
\hfill
\includegraphics[width=3.0in]{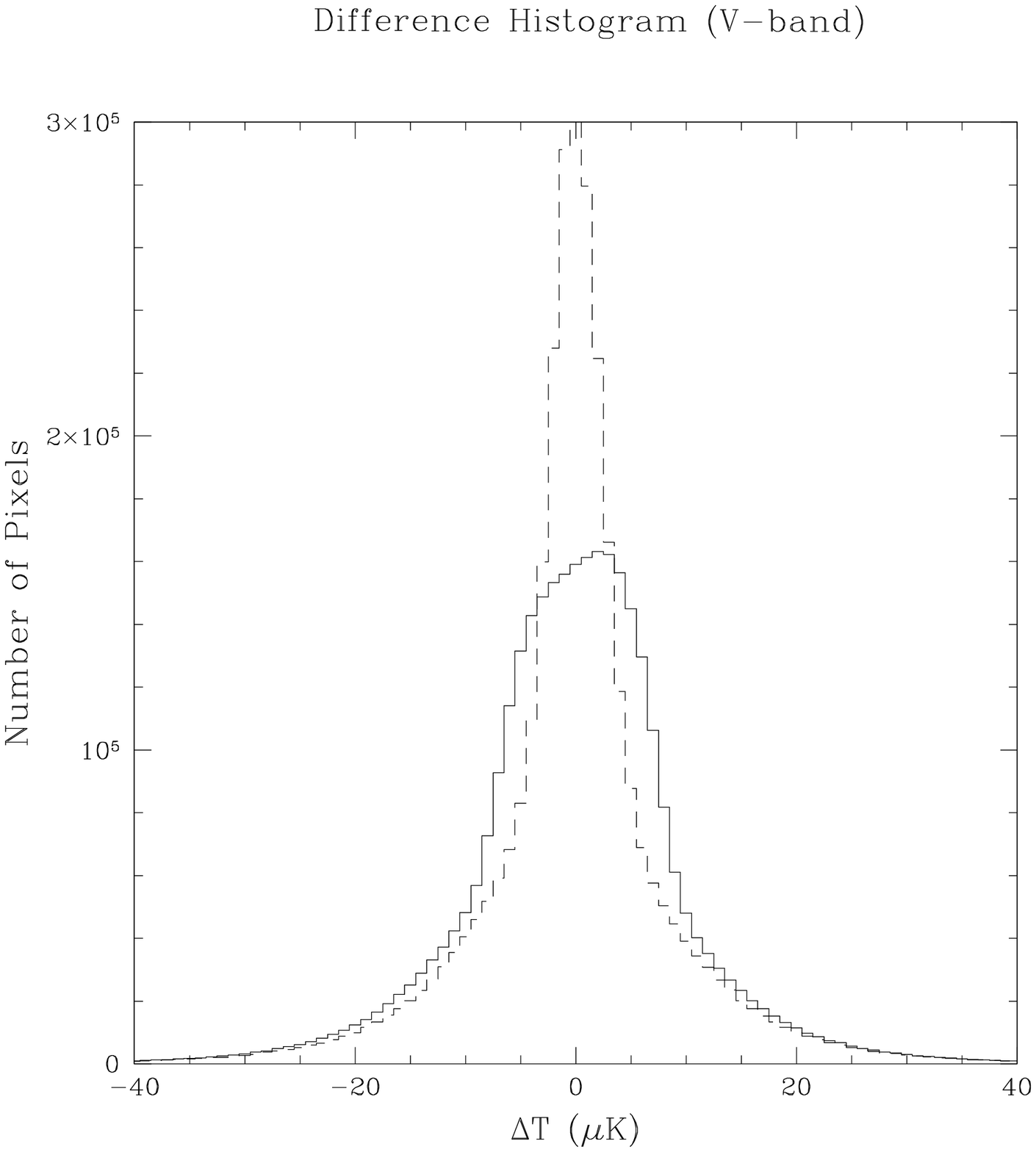}
\hfill
\includegraphics[width=3.0in]{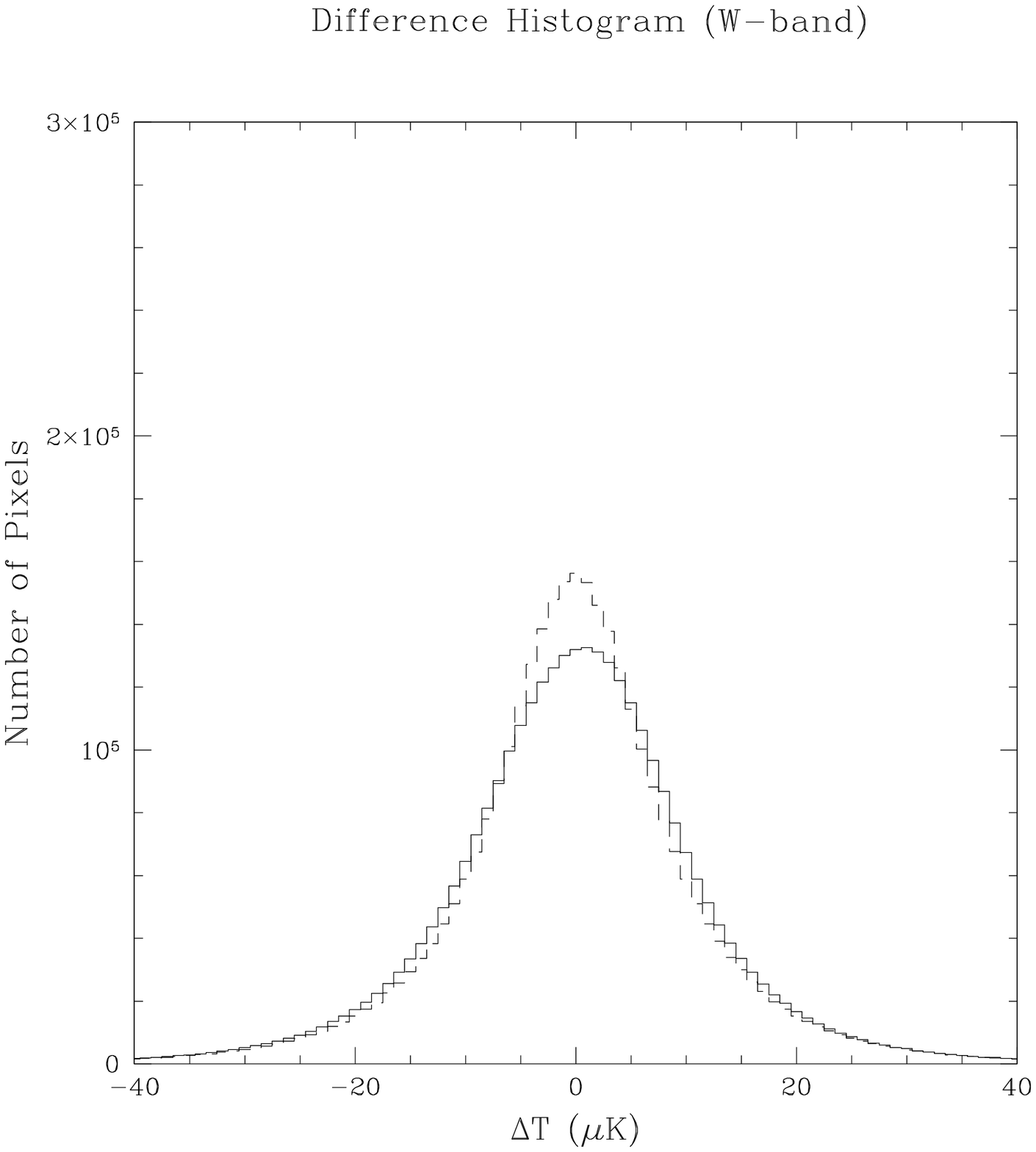}
\caption{Difference histograms ${\Delta}T_{\rm WMAP} - {\Delta}T$ for the
Q-, V-, and W-band data.  The solid
line is a histogram of the data shown in Figure \ref{fig:diffmaps},
while the dashed line corresponds to Figure \ref{fig:diffmapsnodipole}.
}
\label{fig:diffmapshist}
\end{figure}

Histograms of the temperature differences in all pixels are displayed in
Figure \ref{fig:diffmapshist}.  If we examine
only those pixels for which $n_{\rm obs}$ differs by more than 
1\% between our maps and those of {\it WMAP} (presumed to be 
``speckled pixels"), we find that in the
Q band, the fraction of pixels which are speckled pixels
at 10, 20, and 40 $\mu$K is $\approx$ 0.07, 0.25, and 0.70 respectively.
In the V band the respective fractions are 0.08, 0.20, and 0.70, while in the
W band, they are 0.03, 0.10, and 0.40.  Speckled pixels thus dominate the
tails of the histograms.

\section{Determination of Asymmetry in Generated Maps}

\label{sect:powspec}

To determine asymmetry in summed power between 
arbitrarily defined opposite hemispheres, we
follow Eriksen I and Hansen I by principally 
analyzing a combined map consisting of
foreground-corrected V- and W-band maps.  We perform 
foreground corrections using the method of Bennett et al.~(2003b;
hereafter Bennett II),
and combine the maps using equation (24) of Hinshaw II.
To compute cut-sky pseudo-$C_l$ power spectra, we implement the
Master algorithm of Hivon et al.~(2002), summarized in Appendix A
of Hinshaw II.\footnote{
We compute spherical harmonic transforms using the
{\tt ccSHT} library; see\\
{\scriptsize \tt http://crd.lbl.gov/$\sim$cmc/ccSHTlib/doc/}.}
This differs from Eriksen I and Hansen I, who use the maximum
likelihood method of Hansen et al.~(2002) and Hansen \& G\'orski (2003) 
to compute cut-sky power spectra.
As discussed in {\S}2.2 of Efstathiou (2004), use of the Hivon et al.~method is 
not optimal in that there is a estimator-induced variance that makes
it an inaccurate determiner of power in the low-$l$ regime, with the
variance of the sampling distribution increasing with mask size.
In this particular work, however, the use of the Hivon et al.~method 
does not have a detrimental effect in that
we are not interested in power amplitudes or the magnitudes of the
ratio per se, but rather in how the
observed ratios compare with sampling distributions that are determined
via simulations and thus have
the estimator-induced variance built in.

To compute asymmetry as a function of location on the sky, 
we define 101 independent NP orientations
on the northern galactic hemisphere ($l_{\rm gal}$~=~[0$^{\circ}$,324$^{\circ}$] with ${\Delta}l_{\rm gal}$~=~36$^{\circ}$ and
$b_{\rm gal}$~=~[0$^{\circ}$,90$^{\circ}$] with
${\Delta}b_{\rm gal}$~=~9$^{\circ}$; $l_{\rm gal}$ is not defined for 
$b_{\rm gal}$~=~90.0$^{\circ}$).\footnote{
Our grid is not an equal-area grid on the sky (unlike Eriksen I and Hansen I)
and is primarily designed to assist visualization via
contour plots.}  
We apply two Kp2+hemispherical masks for each NP orientation and compute
power spectra for the northern and southern hemispheres for 
$l \leq$~64.  We then calculate asymmetry 
over arbitrary multipole ranges, using arbitrary weightings
(e.g., $w_l$~=~1, or $w_l~=~l(l+1)$), via:
\begin{equation}
A_{\rm obs}(l_{\rm gal},b_{\rm gal},l_{\rm min},l_{\rm max},w_l)~=~\frac{\sum_{l_{\rm min}}^{l_{\rm max}} w_l C_l^{\rm NH}}{\sum_{l_{\rm min}}^{l_{\rm max}} w_l C_l^{\rm SH}} \,.
\end{equation}
(In this work, we assume $w_l~=~l(l+1)$.)
We examine each direction separately because of the number
of pixels within the north and south masks is generally unequal;
this causes the expected mean asymmetry value derived from simulations
to differ from unity and to vary as a function of direction.

We compare our results with simulated sampling distributions.
We use the {\tt HEALPix} facility {\tt synfast} to generate
random universes by sampling from a given power spectrum.
Current theoretical bias leads us to adopt
the LCDM power spectrum derived from a running-index primordial
spectrum using {\it WMAP}, {\it CBI}, and {\it ACBAR} data 
(which we denote \dataset[ADS/Sa.WMAP\#lcdm_ri_model_yr1_v1.txt]{RI}).
We simulate 512 random universes, generating two copies of each.  We convolve
the two copies with the
21$\arcmin$ V-band and 13$\arcmin$ W-band beam responses respectively.
From the two band-specific maps, we generate six
noisy radiometer-specific maps, where we assume Gaussian pixel
noise $\sim N(0,\sigma_o/\sqrt{n_{{\rm obs},p}})$ and use tabulated 
radiometer-specific values of $\sigma_o$.\footnote{
{\scriptsize \tt http://lambda.gsfc.nasa.gov/product/map/pub\_papers/firstyear/supplement/WMAP\_supplement.pdf}}
We co-add and analyze the six noisy simulation 
maps in the manner of our six observed
maps, generating a distribution of 512 asymmetry values in each defined
direction $(l_{\rm gal},b_{\rm gal})$.

\begin{figure}[p]
\centering
\vskip -0.5in
\includegraphics[width=4.0in]{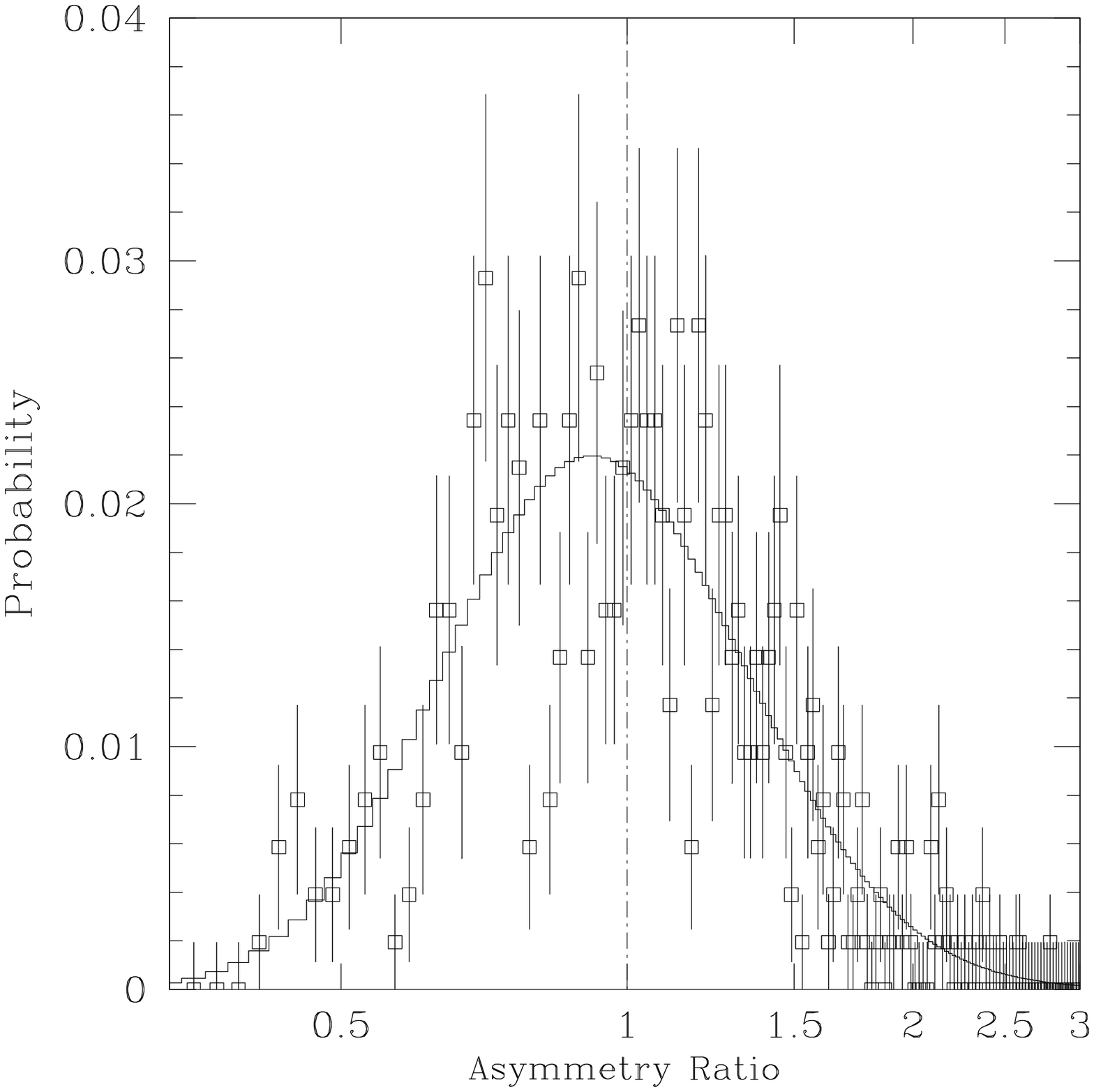}
\hfill
\vskip -0.30in
\includegraphics[width=4.0in]{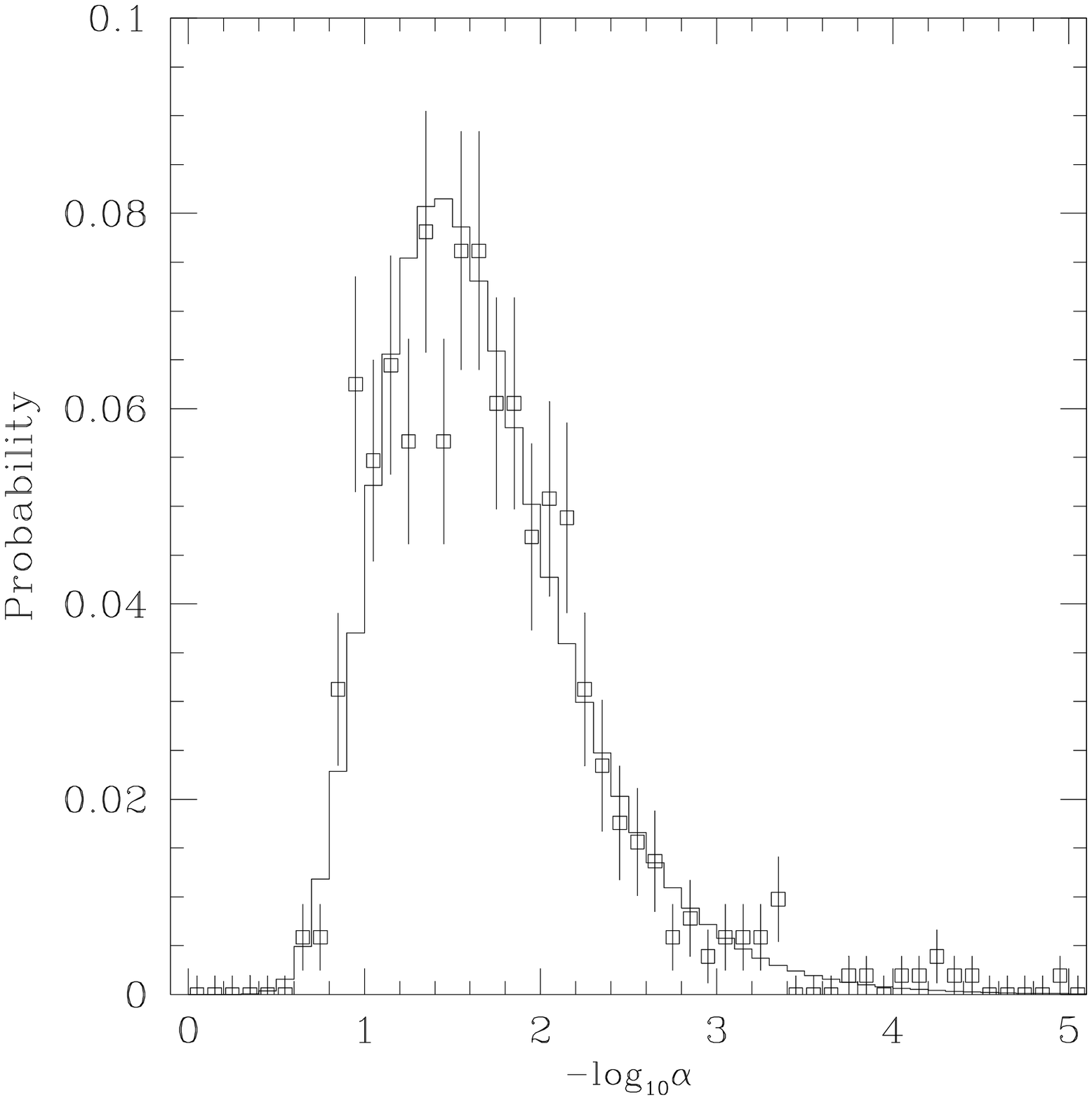}
\vskip -0.20in
\caption{
Example fit of a lognormal distribution to 512 simulated asymmetry values
for given direction $(l_{\rm gal},b_{\rm gal})$,
which is used to estimate $\alpha_{\rm dir}$ (top panel; see equation \ref{eqn:dir}),
and example fit of an extreme-value distribution to
512 simulated values of $\alpha_{\rm dir,min}$, which is used
to estimate $\alpha_{\rm global}$ (bottom panel; see equation \ref{eqn:glob}).
}
\label{fig:dist}
\end{figure}

We observe that the sampling distribution for 
$A(l_{\rm gal},b_{\rm gal})$,
the ratio of two Gaussian-distributed random variables under the 
null hypothesis, is approximately lognormal:
\begin{equation}
A_{\rm obs}(l_{\rm gal},b_{\rm gal})~\sim~f(A,\mu,\sigma)~=~\frac{1}{\sqrt{2\pi}(A-\mu)\sigma} \exp\left[-\left(\frac{\ln(A-\mu)}{\sqrt{2}\sigma}\right)^2\right] \,.
\end{equation}
This is expected because if, for instance, the distribution mean is unity, one
would expect the same probability for observing asymmetry values of, e.g.,
0.5 and 2.0.
We fit lognormal distributions to the data to estimate
significances (see Figure \ref{fig:dist}); otherwise, estimates would be limited
to values $\alpha \gtrsim 1/512 \sim 10^{-3}$ and would be
inaccurate in the distribution tails.
For a given set of parameters
$(l_{\rm gal},b_{\rm gal},l_{\rm min},l_{\rm max})$,
we sum over the multipole range,
obtain best-fit values $\mu_o$ and $\sigma_o$ by
maximizing the sum of the log-likelihoods for each datum,
and determine what we call
the directional significance, $\alpha_{\rm dir}$:
\begin{equation}
\alpha_{\rm dir}~=~{\rm min}[\int_0^{A_{\rm obs}}dA f(A,\mu_o,\sigma_o)~,~\int_{A_{\rm obs}}^{\infty}dA f(A,\mu_o,\sigma_o)] \,.
\label{eqn:dir}
\end{equation}
$\alpha_{\rm dir}$ is an estimate of the probability that a random
Gaussian field exhibits at least as much power asymmetry as observed in 
the data in the specific direction $(l_{\rm gal},b_{\rm gal})$.

While the directional significance is useful for, e.g., visualization, it
is not sufficient to establish the global significance of asymmetry
for a given dataset, $\alpha_{\rm global}$, which
is an estimate of the probability that the random Gaussian field
exhibits at least as much asymmetry as observed in the data in 
{\it any} direction.
We cannot easily obtain $\alpha_{\rm global}$
because the directional distributions
are correlated, with the degree of correlation dependent on
the multipole range considered.
(We observe that because the distributions are correlated,
the simplest Bonferroni correction$-$multiplying
the smallest observed value of directional significance,
$\alpha_{\rm dir,min}^{\rm obs}$, by the number of 
examined directions, 101$-$is too conservative.)  A rigorous approach
would involve transforming our simulation results
to log-space and numerically determining the
correlations between all combinations of grid points, which would allow
us to specify the 101-dimensional Gaussian probability function.  We would
then examine this function using methods described in, e.g., Adler (2000)
to estimate $\alpha_{\rm global}$.
However, as this calculation is very computationally intensive,
we adopt a simpler approach: we compare
$\log(\alpha_{\rm dir,min}^{\rm obs})$
with its probability distribution, which
we find to be approximately an extreme-value distribution:
\begin{equation}
\log(\alpha_{\rm dir,min}^{\rm obs})~\sim~g[\log(\alpha_{\rm dir,min})]~=~\frac{1}{\sigma}\exp\left[\frac{\mu-\log(\alpha_{\rm dir,min})}{\sigma} - \exp\left(\frac{\mu-\log(\alpha_{\rm dir,min})}{\sigma}\right)\right] \,.
\end{equation}
As this distribution is the limiting distribution
for the largest element of a set of independent samples from
a normal distribution, this finding is not unexpected.
In a similar manner to that described above, we determine best-fit
values $\mu_o$ and $\sigma_o$ (see Figure \ref{fig:dist}); 
the estimated global significance is then:
\begin{equation}
\alpha_{\rm global}~=~1 - \exp\left[-\exp\left(\frac{\mu_o-\log(\alpha_{\rm dir,min}^{\rm obs})}{\sigma_o}\right)\right] \,.
\label{eqn:glob}
\end{equation}
We consider $\alpha_{\rm global} \leq$~0.05 to be significant.
We make no claim that this computation is the optimal proxy for
a full Gaussian random field calculation;
more accurate and/or powerful, yet still computationally simple, 
hypothesis tests may exist.

\begin{figure}[p]
\centering
\includegraphics[width=6.0in]{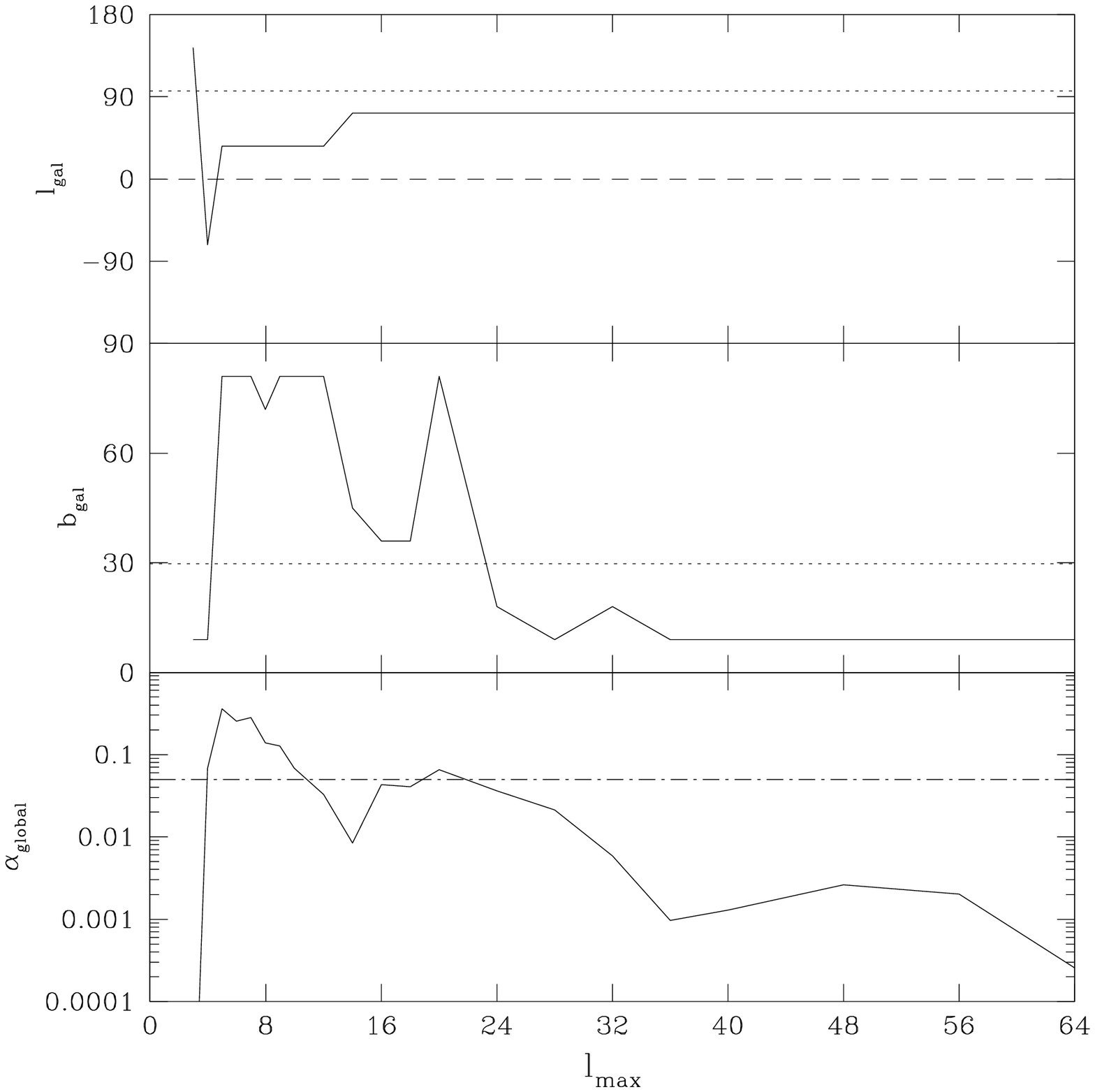}
\caption{
Direction of maximum asymmetry in galactic coordinates (top two panels), 
and the global significance $\alpha_{\rm global}$ (bottom panel),
computed in bins $l$~=~[2,$l_{\rm max}$] for the RI input power spectrum.
The horizontal dotted line in the top two
panels indicates the position of the ecliptic NP, while
the dot-dashed line in the bottom panel marks $\alpha_{\rm global}$~=~0.05.
}
\label{fig:sigrange2}
\end{figure}

In Figure \ref{fig:sigrange2} we show the direction of maximum significance
and $\alpha_{\rm global}$ for the multipole range $l$~=~[2,$l_{\rm max}$].
Unlike Hansen I, who find that the direction of maximum significance
is always toward the galactic NP if $l_{\rm min}$~=~2, we find that it
moves from the vicinity of the galactic NP toward
the galactic plane as $l_{\rm max}$ increases.
For instance, for the range $l$~=~[2,40], we observe that 
$\alpha_{\rm dir}^{\rm plane} \sim 10^{-3}\alpha_{\rm dir}^{\rm NP}$.
For $l_{\rm min} \geq$ 5, we find results consistent with those of Hansen I.
Over the full multipole range that we consider, $l$~=~[2,64],
we find a maximum directional significance 
$\alpha_{\rm dir}$~=~3.3$\times$10$^{-6}$
in the direction $(l_{\rm gal},b_{\rm gal})$~=~(72$^{\circ}$,9$^{\circ}$),
with global significance $\alpha_{\rm global}$~=~2.6$\times$10$^{-4}$.
(This is consistent with the $\sim$ 10$^{-3}$ value determined by
Hansen I using 2048 simulations, without fits of sampling distributions.)
This result indicates clearly that there
is power asymmetry in the {\it WMAP} data.  

\begin{figure}[p]
\centering
\includegraphics[width=6.0in]{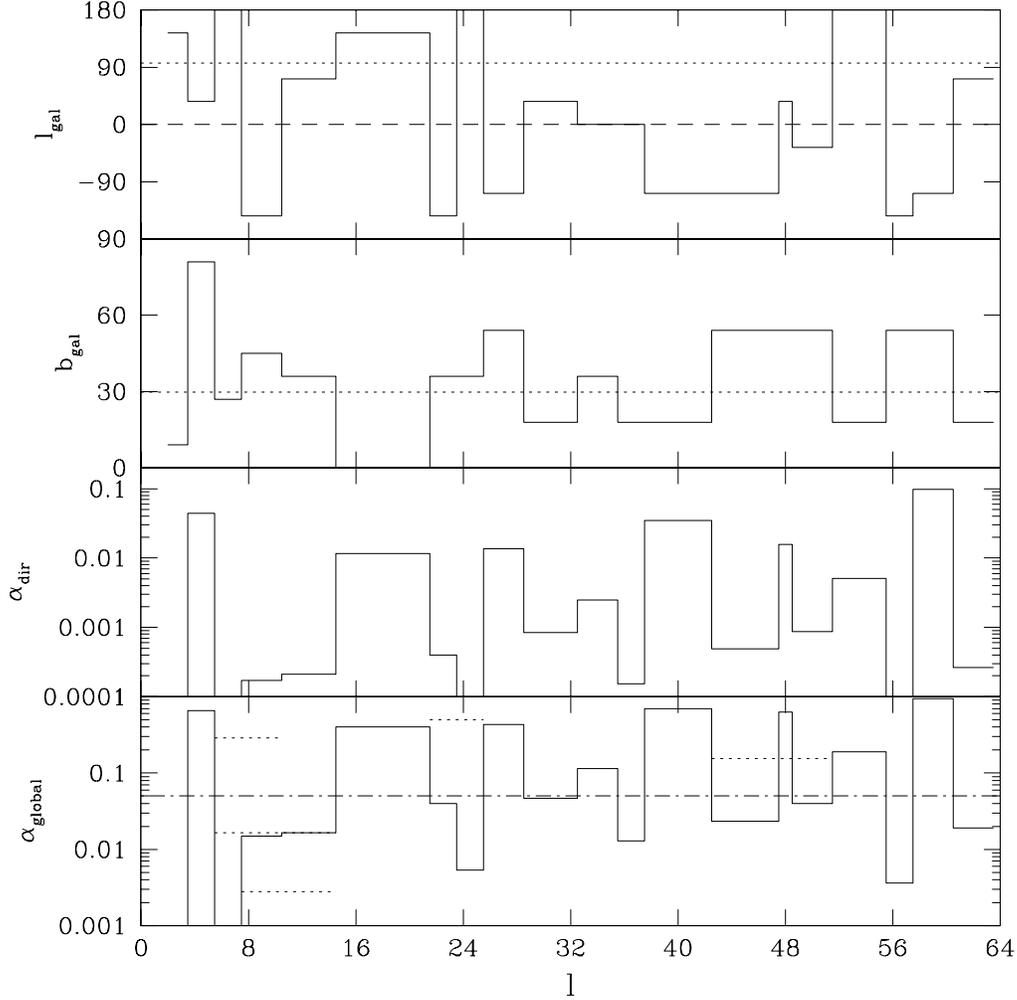}
\caption{
Direction of maximum asymmetry in galactic coordinates (top two panels), 
the corresponding significance $\alpha_{\rm dir}$ (third panel),
and the global significance $\alpha_{\rm global}$ (bottom panel),
as a function of $l$, for the RI input power spectrum.
See {\S}\ref{sect:powspec}
for a description of how bin widths are determined.
The horizontal dotted line in the top two
panels indicates the position of the ecliptic NP, while
the dot-dashed line in the bottom panel marks $\alpha_{\rm global}$~=~0.05.
The additional dotted lines in the bottom panel indicate the
values of $\alpha_{\rm global}$ if we combine (nearly) contiguous
significant ranges: $l$~=~[6,10], [6,14], [8,14], [22,25], and [43,51].
}
\label{fig:asym_ri}
\end{figure}

At what scales is power asymmetry most evident?  Mindful that
there could be a number of causes
manifesting themselves at different scales,
we determine the {\it smallest} ranges with significant asymmetry.
We compute $\alpha_{\rm global}$ in multipole bins with minimum width
${\Delta}l$~=~2.
(We do not examine each multipole singly because the computation
of $\alpha_{\rm global}$ is complicated at low $l$ by the
fact that the sampling distributions for $C_l$ are truncated at
zero, resulting in a considerable number of asymmetry values that
are either zero or infinite.  This is not an issue for ${\Delta}l \geq$ 2.)
We start with ${\Delta}l$~=~2, and
increase the bin size until no more significant bins are found.
For a given ${\Delta}l$, we examine all bins (e.g., for
${\Delta}l$~=~2, we examine $l$~=~[2,3], then [3,4], etc.), and
mark those for which $\alpha_{\rm global} \leq$~0.05.  If two or more
significant bins overlap, we mark only the most significant bin.  For
each ${\Delta}l$, we examine only those multipoles that have not yet
been marked as part of a significant bin (e.g., if we find that only
the bin $l$~=~[2,3] is significant for ${\Delta}l$~=~2, the first
bin we would examine for ${\Delta}l$~=~3 is [4,6], and not [2,4]).
We note that this binning algorithm is not meant as a rigorous 
statistical exercise (we examine so many bin combinations that
a number of of the chosen bins may be false positives) but rather
as a guide to further examination of the data.
In Figure \ref{fig:asym_ri}
we show the result: we find twelve significant ranges, including two
pairs and one triplet of (nearly) contiguous bins 
(separated by at most one bin). 
The gaps between significant bins are treated as single,
insignificant bins for purposes of plotting.
As shown in the bottom panel of Figure \ref{fig:asym_ri},
combining (nearly) contiguous bins does not necessarily yield a 
more significant result.
The large number of significant bins and the fact that 
we discern no general trend in the direction of $A_{\rm obs}^{\rm max}$
lead us to conclude that the process(es) contributing
to the observed asymmetry (e.g., residual foreground emission) 
are spatially complex.

The bins $l$~=~[2,3] and $l$~=~[6,7] appear anomalously significant, with
$\alpha_{\rm global}~\lesssim$~10$^{-6}$ (the value depends 
very sensitively upon the fit parameters of the extreme-value
distribution and we cannot determine it accurately).
For $l$~=~[2,3], the direction of maximum significance is
($l_{\rm gal},b_{\rm gal}$)~=~(144$^{\circ}$,9$^{\circ}$), with
$C_{2,3}^N$~=~\{9.7,1.1\} and $C_{2,3}^S$~=~\{301.7,755.9\}, while
for $l$~=~[6,7], the corresponding values are ($l_{\rm gal},b_{\rm gal}$)~=~(180$^{\circ}$,27$^{\circ}$), $C_{6,7}^N$~=~\{303.6,196.7\}, 
and $C_{6,7}^S$~=~\{51.7,0.0\}.
We confirm that the simulated values of $\alpha_{\rm global}$
in each case follow extreme-value distributions with no discernible
deviation that may affect the estimation of significance.
Thus we take the observed significances at face value and conclude that
there is highly significant power asymmetry at the largest scales.
We note that we determine the direction of maximum asymmetry for $l$~=~[2,3] 
to be in the galactic plane, while for Hansen I it is the galactic NP 
(for $l$~=~[2,4]; we confirm that our result does not change if we include
$l$~=~4).

To determine the sensitivity of
$\alpha_{\rm global}$ to changes in frequency, we simulate
separate sets of co-added Q-, V-, and W-band data, as well as a set
of co-added data from all three bands.  In all cases, we find
results that are similar, both qualitatively and quantitatively, to 
those presented above; for instance, over the multipole range $l$~=~[2,64], we
find $\alpha_{\rm global}$~=~3.0$\times$10$^{-4}$ (Q),
6.2$\times$10$^{-4}$ (V), 7.1$\times$10$^{-4}$ (W), and
2.3$\times$10$^{-4}$ (all bands).  This finding is consistent with
that of Hansen I, who analyze Q-band data with a Kp0 mask and find
persistent asymmetry.

\begin{figure}[p]
\centering
\includegraphics[width=6.0in]{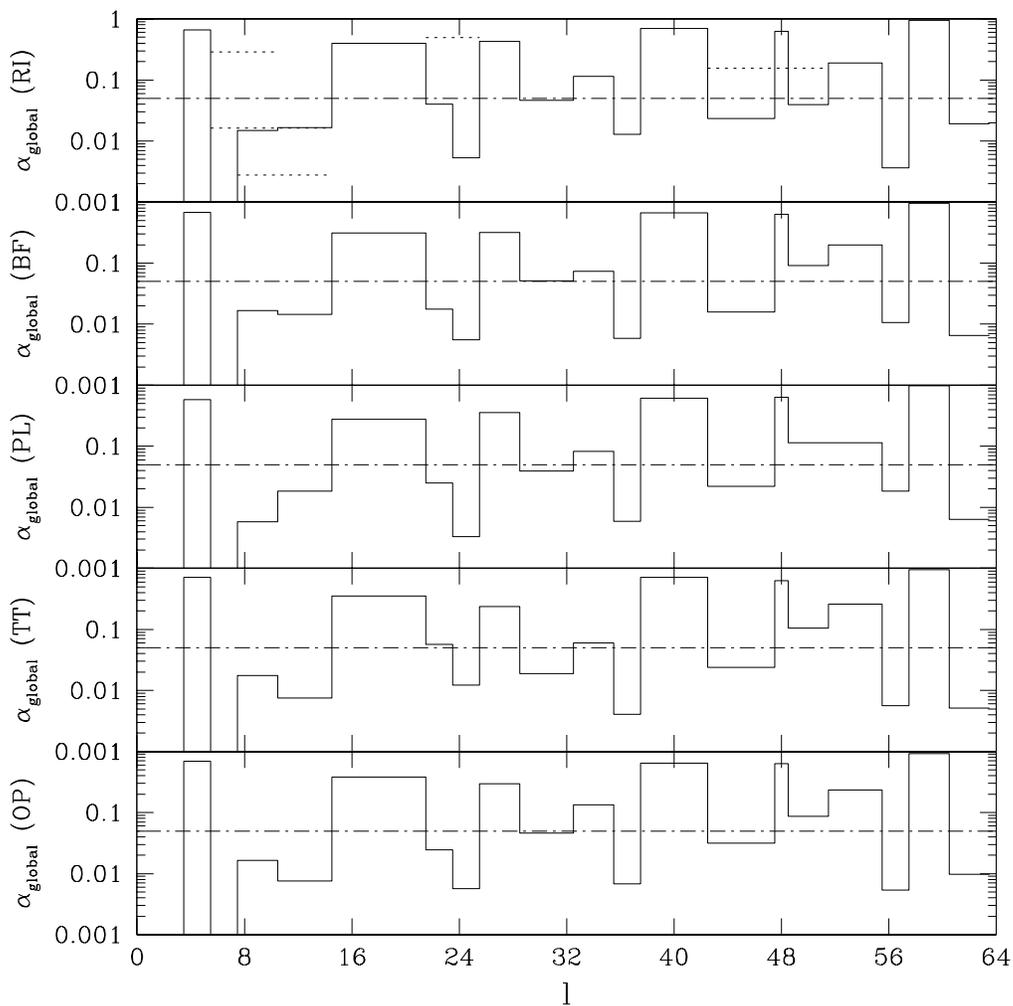}
\caption{
Similar to the bottom panel of Figure \ref{fig:asym_ri},
showing global significance $\alpha_{\rm global}$ as a function
of $l$ for the five different input power 
spectra listed in the text.
The top panel of this figure is the same as the bottom panel of
Figure \ref{fig:asym_ri}.
This figure demonstrates that $\alpha_{\rm global}$
is relatively insensitive to the choice of power 
spectrum.
}
\label{fig:asym_globcmp}
\end{figure}

We also repeat our analysis using four alternate 
theoretical\footnote{\scriptsize
{\tt http://lambda.gsfc.nasa.gov/product/map/wmap\_lcdm\_models.cfm}}
and phenomenological power spectra as input to simulations:
\begin{enumerate}
\item a running-index spectrum similar to the RI spectrum, 
with {\it 2dF} and Lyman-$\alpha$ data also
included in the computation (\dataset[ADS/Sa.WMAP\#lcdm_bf_model_yr1_v1.txt]{BF});
\item the theoretical LCDM power spectrum derived from a power-law primordial 
spectrum (\dataset[ADS/Sa.WMAP\#lcdm_pl_model_yr1_v1.txt]{PL}); 
\item the observed {\it WMAP} ``One-Year Combined TT Power Spectrum'' 
(\dataset[ADS/Sa.WMAP\#comb_tt_powspec_yr1_v1p1.txt]{TT});\footnote{\scriptsize
{\tt http://lambda.gsfc.nasa.gov/data/map/powspec/map\_comb\_tt\_powspec\_yr1\_v1p1.txt}}
\item and the same as immediately above, but optimally smoothed via
local linear smoothing, with bandwidth chosen by leave-one-out 
cross-validation (see, e.g., Fan \& Gijbels 1996; OP).
\end{enumerate}
For each spectrum, we compute $\alpha_{\rm global}$ within the previously
defined RI bins, as opposed to defining a new set of bins.  This has
little effect on our results, which we display in Figure \ref{fig:asym_globcmp}.
We find that $\alpha_{\rm global}$ is largely insensitive to the choice of
power spectrum; while the variance of the distribution of 
simulated asymmetry values differs for each, the differences are
insufficient to qualitatively affect results.  However, we do note
that the range $l$~=~[49,51], significant given the RI spectrum, is
not significant for any other spectrum.

\begin{figure}[p]
\figurenum{9a}
\plotone{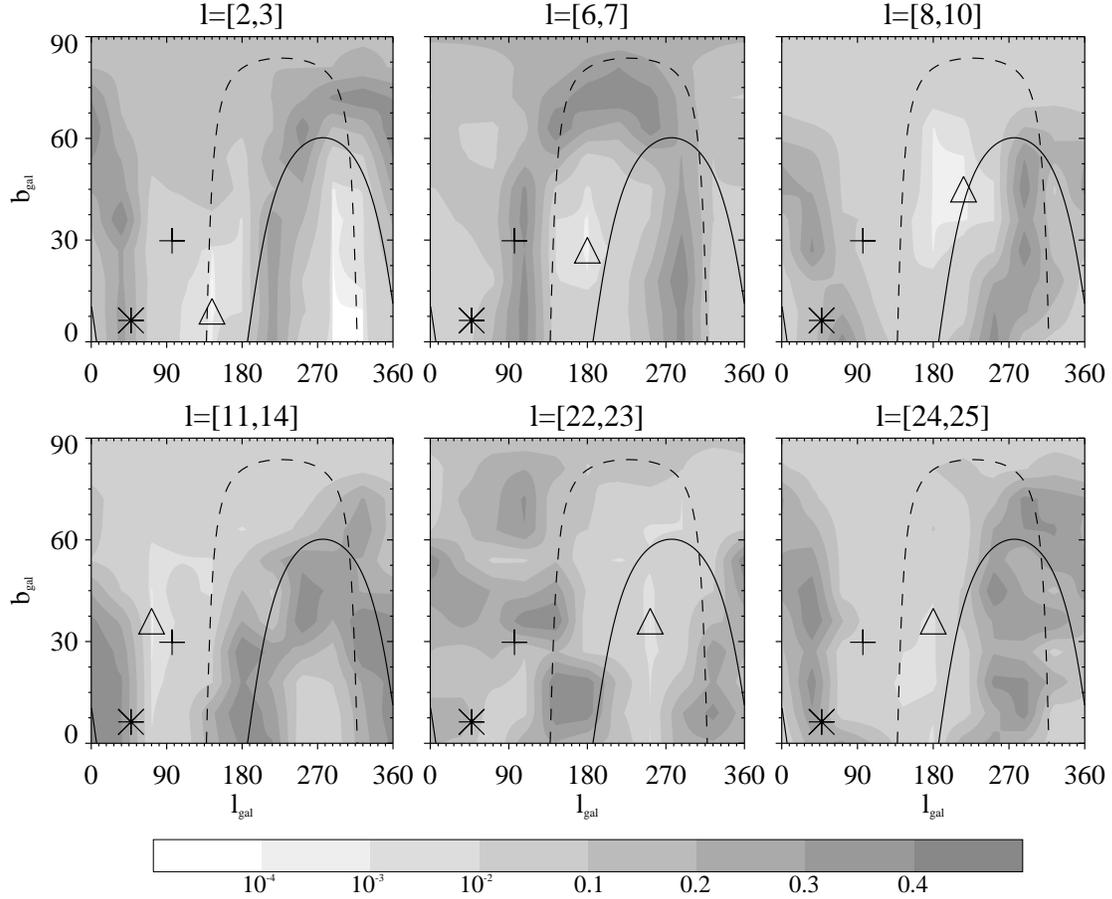}
\caption{
Directional significance contour plots
showing the significance of observed asymmetry
as a function of NP location in $(l_{\rm gal},b_{\rm gal})$
within multipole bins
in which the global significance of
asymmetry is $\alpha \leq$~0.05.  The minimum bin width is 
${\Delta}l$~=~2.
The direction of maximum significance 
is labeled with a triangle, the cross and solid line
indicate the ecliptic NP and plane respectively,
and the asterisk and the dashed line indicate the
supergalactic NP and plane respectively.
}
\label{fig:c12-1}
\end{figure}

\begin{figure}[p]
\figurenum{b}
\plotone{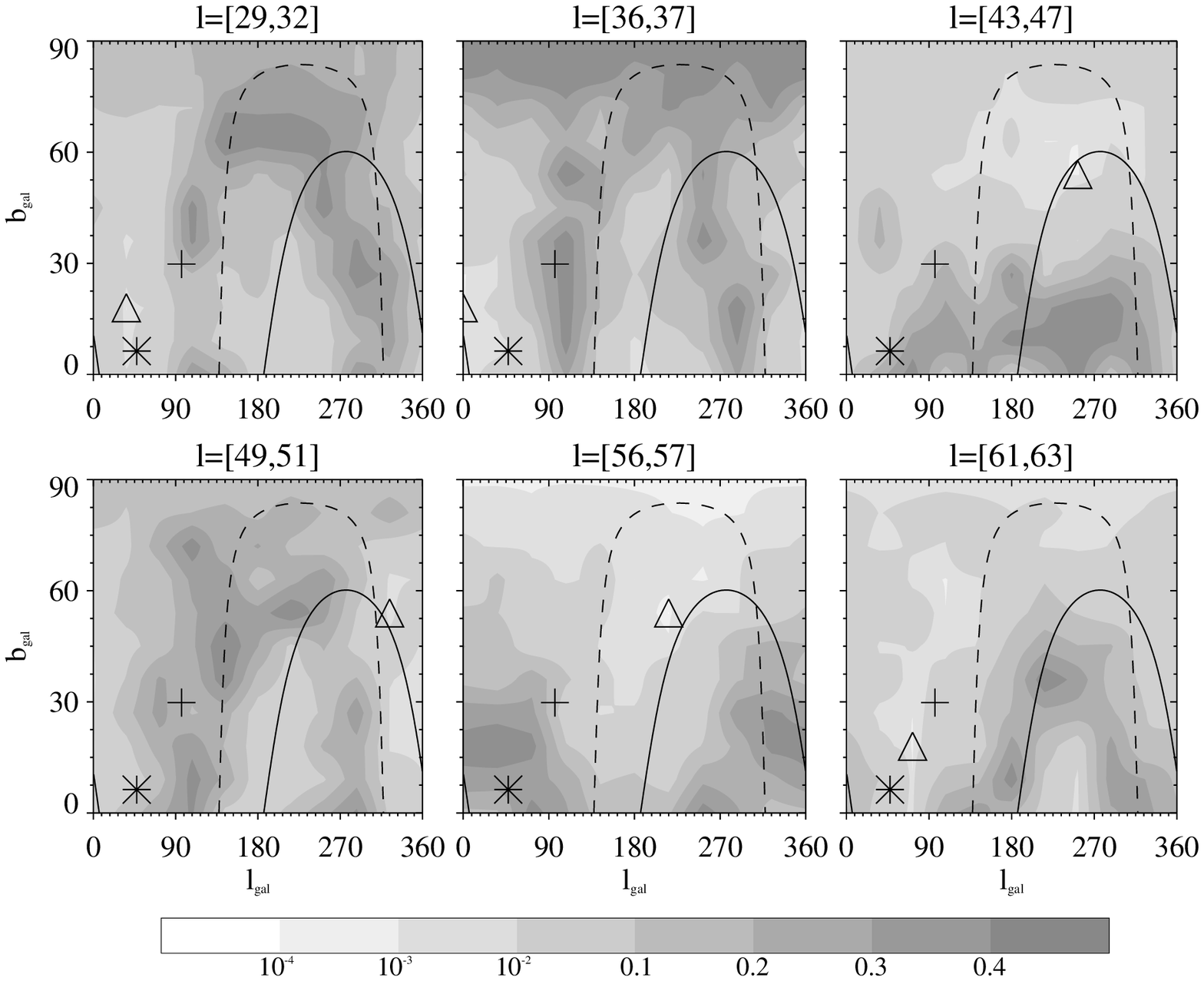}
\raggedright
Fig. 9b.
\label{fig:c12-2}
\end{figure}

In Figures \ref{fig:c12-1}-\ref{fig:c12-2}, we display
contours of $\alpha_{\rm dir}$ for each of
the twelve globally significant multipole ranges shown in Figure
\ref{fig:asym_ri}.  The direction and
significance of maximum asymmetry in each panel in these figures
maps directly
back to the values plotted in the top three panels of Figure \ref{fig:asym_ri}.
Interestingly, these contours exhibit evidence of a relationship between the
ecliptic plane and both the direction of maximum significance and
the contours of {\it least} significance ($\alpha_{\rm dir} \approx$ 0.2-0.5).
We find that in four of the twelve plots, the direction of maximum
asymmetry lies within $\approx$~4$^{\circ}$ of the ecliptic plane and
in eight of twelve, within $\approx$~20$^{\circ}$.  The coarseness of
our coordinate grid precludes an exact quantitative interpretation, 
but we note that the probability of sampling a point within 4$^{\circ}$
of the ecliptic plane
is 0.07, and within 20$^{\circ}$, 0.34; the probability of what we
observe is $p_{\rm binomial} \lesssim$ 0.01 in both cases.

We also observe that, for large scales,
the contours of least significance
appear to follow great circles; furthermore, the great circle for
$l$~=~[6,7] passes over the ecliptic poles.
(Great circle-like structures appear at other scales as well,
but not as strikingly.)
While the significance of such an alignment of contours is difficult to
quantify, we visually examine contour plots created from 
simulated datasets and find no discernible trend
toward the formation of great circles.  
In Figure \ref{fig:gc}, we concentrate on the lowest four multipole
ranges shown in Figure 8.  The dotted and dash-dotted lines in these
figures represent great circles inclined relative to 
the ecliptic and galactic planes, respectively.  
Note that we place the lines by eye and make a purely qualitative
comparison of goodness-of-fit.  For $l$~=~[2,3] and $l$~=~[6,7],
great circles defined in ecliptic coordinates with inclinations
$\approx$ 30$^{\circ}$ and $\approx$ 90$^{\circ}$ respectively appear to fit
the contours of least significance
better than their galactic coordinate counterparts;
for $l$~=~[8,10] and [11,14], there is no clear preference for either
coordinate system.  
In Figure \ref{fig:gcprime}, we demonstrate that
the contours and great circles for $l$~=~[2,3] 
are frequency independent: they are observed
within each radiometer band separately, with the results in each band 
relying on completely independent sets of simulations.
We conclude that whatever causes power asymmetry at the largest
scales may be related to the ecliptic,
but at the same time the underlying process is frequency independent.

In Figure \ref{fig:context}, we show the relationship between the
great circle for the $l$~=~[2,3] and
the axes defined by de Oliveira-Costa et al.~that 
maximize the angular momentum dispersion of the quadrupole and octopole
in their wave-function paradigm, as well as the 
four normals to the planes defined by the multipole vectors of
Schwarz et al.  All except the 
${\bf w}^{(3,3,1)}$ vector of Schwarz et al.~point in the same
general direction (the ``axis of evil," as dubbed by Land \& Magueijo 2005a), 
near a contour of minimum significance
($\alpha_{\rm dir} >$ 0.4). 
However, they also point near the CMB dipole.
In the next section, we examine the effect of changing the magnitude of the
velocity of the CMB dipole on observed asymmetry.

\setcounter{figure}{9}

\begin{figure}[p]
\centering
\includegraphics[width=6.5in]{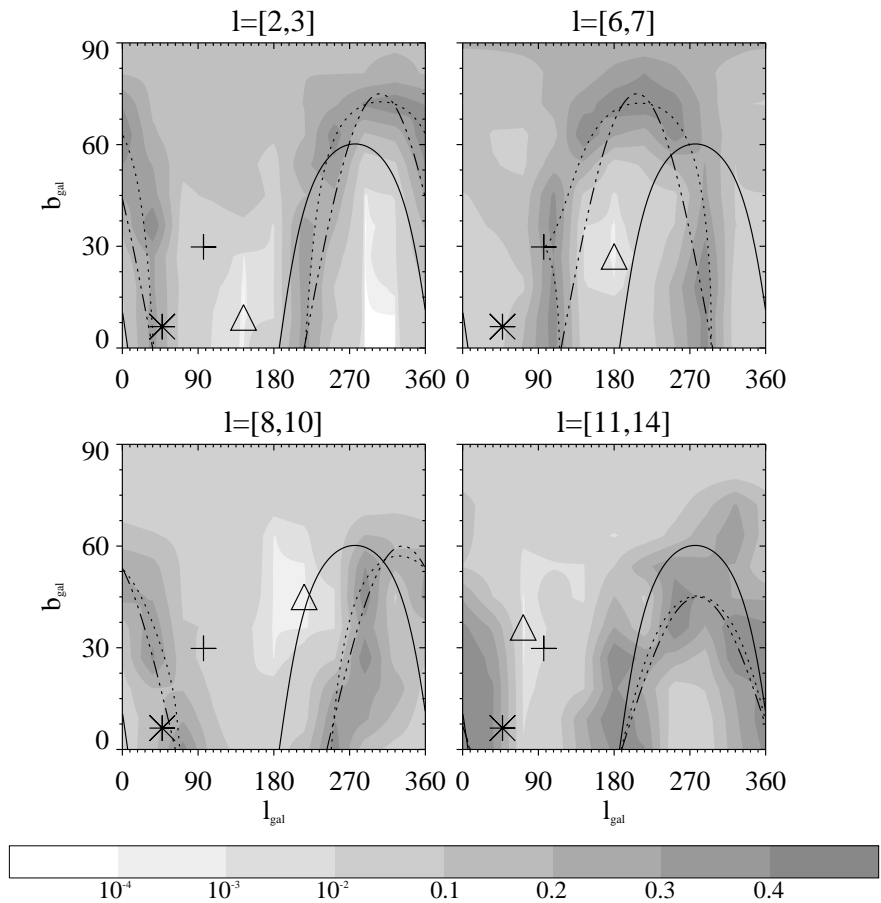}
\caption{
Directional significance contour plots for the lowest four multipole
bins shown in Figure \ref{fig:c12-1} (see this figure for an
explanation of symbols).  All four plots exhibit contours of least
significance that appear to follow great circles.
The dotted and dash-dotted lines represent great circles
inclined relative to the ecliptic and galactic planes, respectively.
The lines are illustrative and their placement is done by eye.
}
\label{fig:gc}
\end{figure}

\begin{figure}[p]
\centering
\includegraphics[width=6.5in]{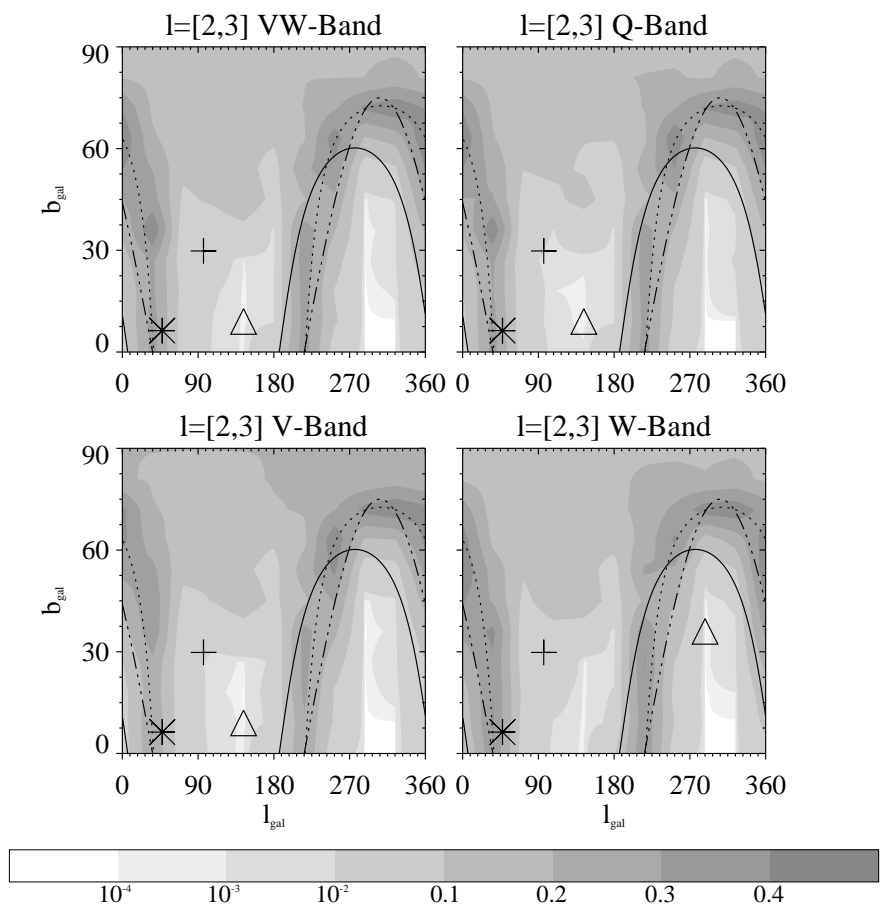}
\caption{
Same as the upper left panel of Figure \ref{fig:gc}, with
co-added Q-, V-, and W-band data displayed in the upper right, lower left, 
and lower right panels respectively.
See Figures \ref{fig:c12-1} and \ref{fig:gc} for explanations of symbols.
We conclude that the process causing power asymmetry at the largest scales
is insensitive to frequency.
}
\label{fig:gcprime}
\end{figure}

\begin{figure}[p]
\centering
\includegraphics[width=6.0in]{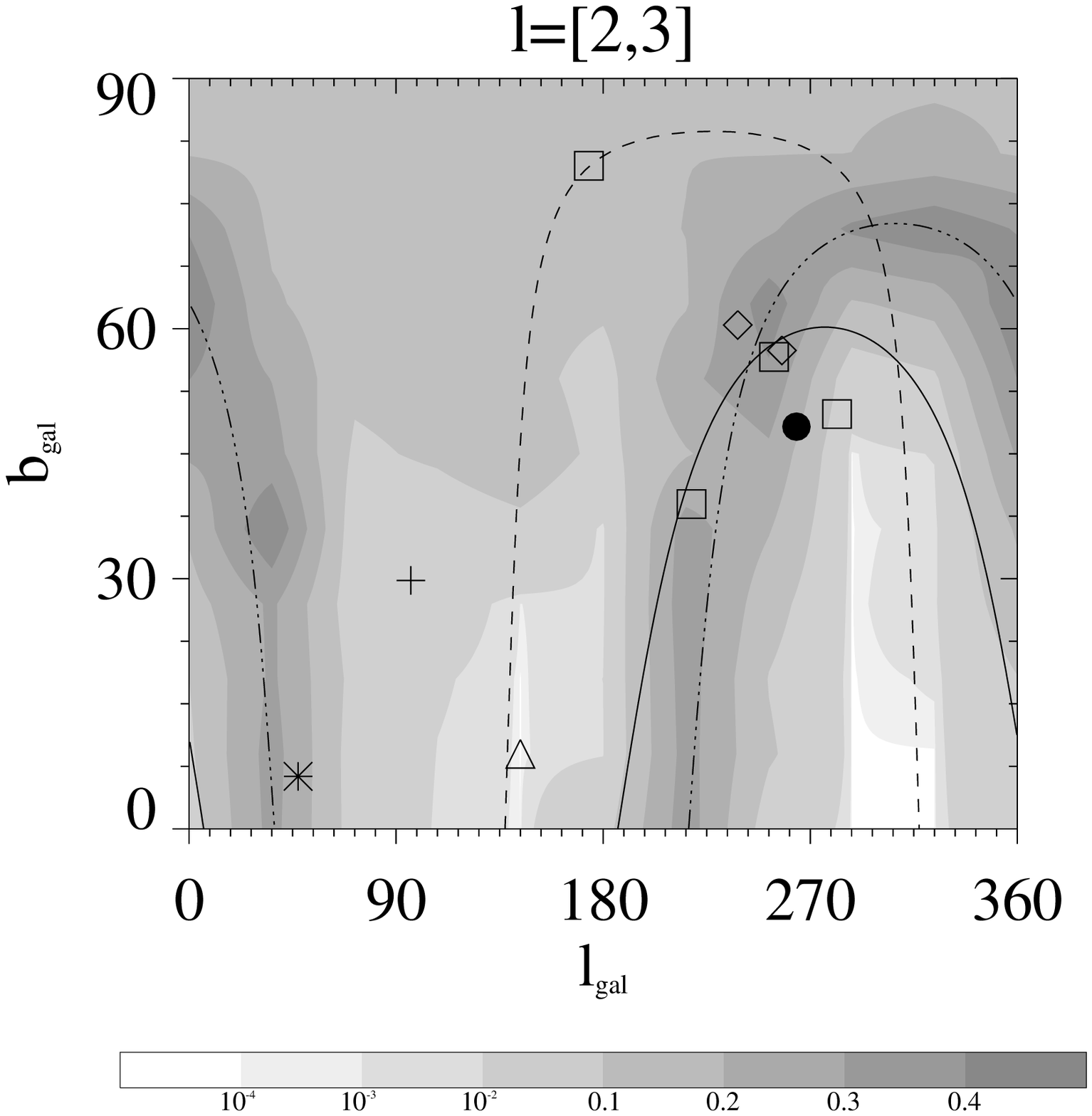}
\caption{
Upper left panel of Figure \ref{fig:gc} with the
great circle in galactic coordinates removed.  
The diamonds represent the pointing direction of the 
quadrupole (lower right) and octopole (upper left) preferred axes defined by 
de Oliveira-Costa et al.,
while the squares represent the pointing direction
of normals to the planes defined by multipole vectors derived by Schwarz et al.
The filled circle shows the direction of the CMB dipole.
}
\label{fig:context}
\end{figure}

\section{Effect of Map-Making Algorithmic Assumptions Upon Observed Asymmetry}

\label{sect:mapassume}

A heretofore unexamined aspect of the asymmetry issue
is the effect of altering the map-making algorithm.  For instance,
what if maps were made using no foreground mask at all, or a reduced
mask, rather than the 
Kp8 mask with no edge smoothing?  Because of the elliptical beam response,
maps would be affected by propagated echoes of bright Galactic sources,
as noted by Hinshaw I.  But would these echoes have any discernible effect on
the observed asymmetry?  (We realize that in this example
using no foreground mask at all is clearly a wrong step no one would
make; our goal in these tests is simply to establish how sensitive
asymmetry is to changes in the map-making algorithm.)

The alterable aspects of our map-making algorithm, both major and minor, 
fall into a number of categories:
\begin{enumerate}
\item The mapping paradigm, involving issues which are too complex to 
      address in the current work:
  \begin{itemize}
  \item[a.] The assumption of Gaussian pixel noise and its subsequent use
        in least-squares estimates.
  \item[b.] The assumption that the matrix product 
        (${\bf M}^{\rm T}{\bf M})^{-1}$ is diagonally dominant and is
        $\approx {\vec n}_{\rm obs}^{-1}$.  This goes hand-in-hand
        with the assumption that the radiometers have $\delta$-function spatial
        response, an assumption which we expect has little to no effect
        in the low-$l$ regime.
  \end{itemize}
\item Aspects which are currently not alterable given the data we have:
  \begin{itemize}
  \item[a.] The algorithm by which the raw data are calibrated (Hinshaw I, {\S}2.2).
  \item[b.] The assumption of equal statistical weights for each datum.
  \end{itemize}
\item Aspects which we directly examine:
  \begin{itemize}
  \item[a.] The assumption that $T_{\rm o}$~=~2.725 K.
  \item[b.] The assumption that the Sun's velocity relative to the CMB is 
            $(v_{\odot,\rm gal}^x,v_{\odot,\rm gal}^y,v_{\odot,\rm gal}^z)$~=\\
            ($-$26.26,$-$243.71,274.63) km s$^{-1}$ (or $v_{\odot,\rm gal}$~=~368.11 km s$^{-1}$).
  \item[c.] The effect of including the loss-imbalance parameters 
            (Table 3 of Jarosik et al.).
  \item[d.] The use of the Kp8 galactic mask without edge smoothing.
  \item[e.] The use of Lagrange interpolating polynomials to determine
            radiometer normal vectors as a function of time.
  \item[f.] The assumption that the planetary cut is 
            $\theta_{\rm cut}$~=~1.5$^{\circ}$.
  \item[g.] A step beyond map-making: the use of the same foreground maps, in
            the same proportion, as
            the {\it WMAP} team (Bennett II, {\S}6).
  \end{itemize}
\end{enumerate}
We also examine the sensitivity of asymmetry to temperature map zero
points, which the {\it WMAP} team derives by fitting a phenomenological
model to southern galactic hemisphere data (Bennett II, {\S}5).  
Normally, changing the zero point, i.e., changing the monopole
component of the anisotropy map, should have no effect on higher
multipoles.  However, power
leakage from the monopole in, e.g., the Master algorithm
means that any uncertainty in the zero point can in theory affect 
measurements of asymmetry.  The average temperatures in the V-band
and W-band maps are $\approx$ 78 $\mu$K and 75 $\mu$K, respectively,
with variance $\sim$ 1 $\mu$K.  We confirm that changing the zero point
by amounts up to 10 $\mu$K has negligible impact upon $\alpha_{\rm global}$.

After altering the value of a given parameter (while holding 
the other parameter values to their default values), we create
new V- and W-band radiometer maps.  We combine these maps into a new
co-added map that we analyze using the methods described
in {\S}\ref{sect:powspec}.  To determine if altering mapping
parameters can reduce or eliminate all asymmetry, we
concentrate upon the range $l$~=~[2,64] (Table \ref{tab:map2to64ri}); 
however, we also examine the range $l$~=~[2,3]
to gauge the effect of changes on large-scale asymmetry.

\begin{deluxetable}{lcccc}
\rotate
\tablenum{1}
\tablecaption{Effect of Changing Mapping Parameters on Power Asymmetry for l~=~[2,64]}
\tablewidth{0pt}
\tablehead{
\colhead{Changed Parameter} & \colhead{$\alpha_{\rm dir}$ Quartiles} & \colhead{$\alpha_{\rm dir,min}$} & \colhead{$(l_{\rm gal},b_{\rm gal})$} & \colhead{$\alpha_{\rm global}$}
}
\startdata
Default Map & (0.152,0.0132,2.56$\times$10$^{-3}$) & 3.30$\times$10$^{-6}$ & (72$^{\circ}$,9$^{\circ}$) & 2.55$\times$10$^{-4}$ \\
No First Day Data & (0.152,0.0132,2.58$\times$10$^{-3}$) & 3.30$\times$10$^{-6}$ & (72$^{\circ}$,9$^{\circ}$) & 2.55$\times$10$^{-4}$ \\
T$_{\rm o}$ = 2.724 K ($-$1$\sigma$) & (0.147,0.013,2.58$\times$10$^{-3}$) & 3.30$\times$10$^{-6}$ & (72$^{\circ}$,9$^{\circ}$) & 2.55$\times$10$^{-4}$ \\
T$_{\rm o}$ = 2.726 K (+1$\sigma$) & (0.152,0.013,2.58$\times$10$^{-3}$) & 3.30$\times$10$^{-6}$ & (72$^{\circ}$,9$^{\circ}$) & 2.55$\times$10$^{-4}$ \\
V$_{\rm d}$ = 366.24 km s$^{-1}$ ($-$1$\sigma$) & (0.144,0.013,2.26$\times$10$^{-3}$) & 4.59$\times$10$^{-6}$ & (72$^{\circ}$,9$^{\circ}$) & 3.46$\times$10$^{-4}$ \\
V$_{\rm d}$ = 369.98 km s$^{-1}$ (+1$\sigma$) & (0.147,0.014,2.75$\times$10$^{-3}$) & 2.35$\times$10$^{-6}$ & (72$^{\circ}$,9$^{\circ}$) & 1.88$\times$10$^{-4}$ \\
V$_{\rm d}$ = 371.85 km s$^{-1}$ (+2$\sigma$) & (0.157,0.017,3.77$\times$10$^{-3}$) & 2.10$\times$10$^{-6}$ & (72$^{\circ}$,9$^{\circ}$) & 1.69$\times$10$^{-4}$ \\
V$_{\rm d}$ = 373.72 km s$^{-1}$ (+3$\sigma$) & (0.147,0.024,5.07$\times$10$^{-3}$) & 2.63$\times$10$^{-6}$ & (72$^{\circ}$,9$^{\circ}$) & 2.08$\times$10$^{-4}$ \\
No Mask & (0.148,0.013,2.86$\times$10$^{-3}$) & 4.11$\times$10$^{-6}$ & (72$^{\circ}$,9$^{\circ}$) & 3.13$\times$10$^{-4}$ \\
No Interpolation & (0.093,7.20$\times$10$^{-3}$,4.39$\times$10$^{-4}$) & 5.70$\times$10$^{-8}$ & (72$^{\circ}$,9$^{\circ}$) & 6.20$\times$10$^{-6}$ \\
Linear Interpolation & (0.152,0.0132,2.58$\times$10$^{-3}$) & 3.30$\times$10$^{-6}$ & (72$^{\circ}$,9$^{\circ}$) & 2.55$\times$10$^{-4}$ \\
Planet Cut: $\theta_{\rm cut}$ = 0.0$^{\circ}$ & (1.45$\times$10$^{-7}$,5.77$\times$10$^{-14}$,2.17$\times$10$^{-19}$) & 1.40$\times$10$^{-26}$ & (72$^{\circ}$,9$^{\circ}$) & 0.0 \\
Planet Cut: $\theta_{\rm cut}$ = 0.5$^{\circ}$ & (0.147,0.0154,2.28$\times$10$^{-3}$) & 2.63$\times$10$^{-6}$ & (72$^{\circ}$,9$^{\circ}$) & 2.08$\times$10$^{-4}$ \\
\enddata
\label{tab:map2to64ri}
\end{deluxetable}

\paragraph{Monopole Temperature.} We find that altering the monopole temperature by 1$\sigma$
(0.001 K) has little effect on the distribution 
of significances, and none on $\alpha_{\rm global}$.

\begin{figure}[p]
\centering
\includegraphics[width=6.0in]{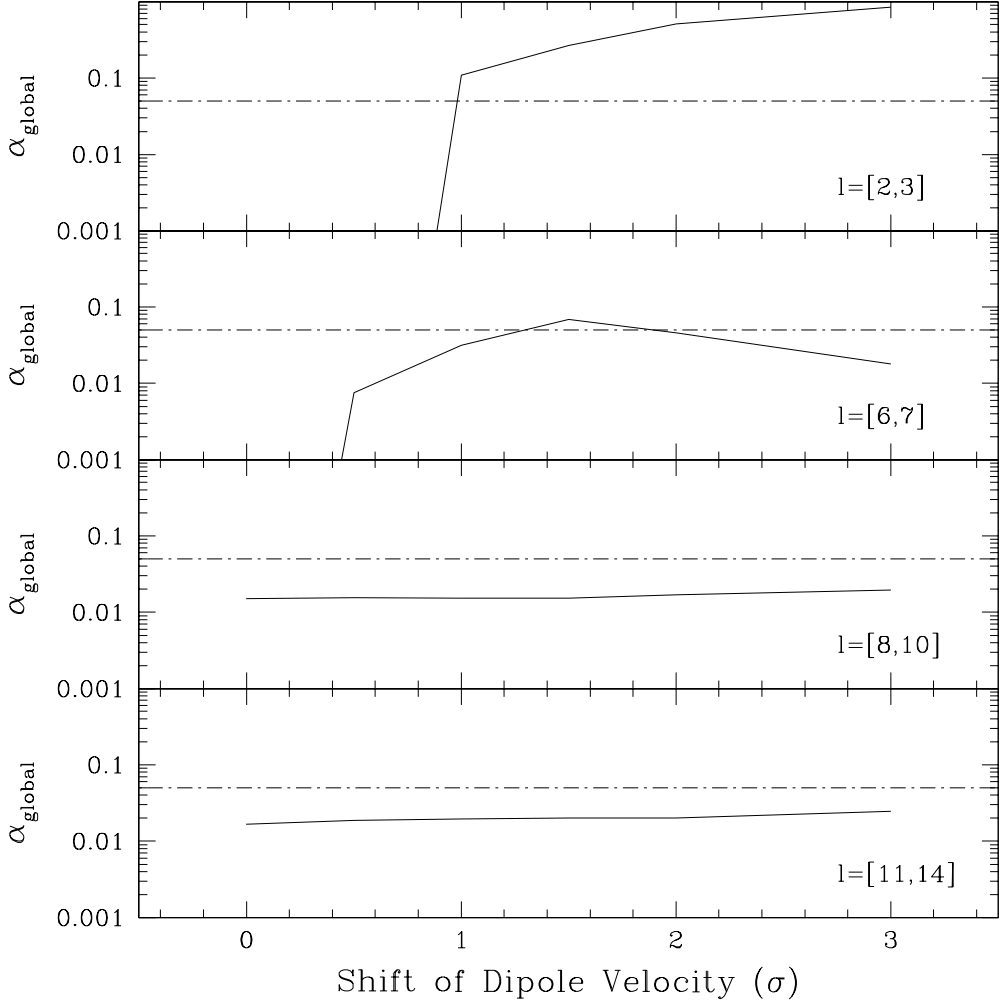}
\caption{
Plots of $\alpha_{\rm global}$ as a function of shifted dipole velocity
(1$\sigma$~=~1.87 km s$^{-1}$), for
the multipole ranges displayed in each panel.  The dot-dashed line in
each panel represents $\alpha_{\rm global}$~=~0.05.  This figure
illustrates that changing the
dipole velocity can have a marked effect upon 
the significance of asymmetry at the
largest scales.  However, since the dipole velocities used to generate
this figure differ from the dipole velocity used to calibrate the
data, the uncertainty of each datum is unknown.
}
\label{fig:vdipole}
\end{figure}

\paragraph{Dipole Velocity Magnitude.}
We test changing the magnitude of the dipole velocity along the direction
($l_{\rm gal}$,$b_{\rm gal}$)~=~(263.85$^{\circ}$,48.25$^{\circ}$).
(Testing changes on a grid of dipole locations is currently too 
computationally intensive.)
We find that positive changes $\gtrsim$~1$\sigma$ ($\gtrsim$~2 km s$^{-1}$)
can have a marked effect upon $\alpha_{\rm global}$ at the largest scales,
even sometimes rendering asymmetry insignificant (Figure \ref{fig:vdipole}).
At smaller scales, the effect on asymmetry is minimal
(e.g., bottom panels of Figure \ref{fig:vdipole}).
Note that the velocities that we assume in map-making do not
match that used to calibrate the data (the {\it COBE} DMR dipole),
so we cannot state with certainty that changing the dipole velocity
can by itself eliminate asymmetry at the largest scales.

\begin{figure}[p]
\centering
\includegraphics[width=6.0in]{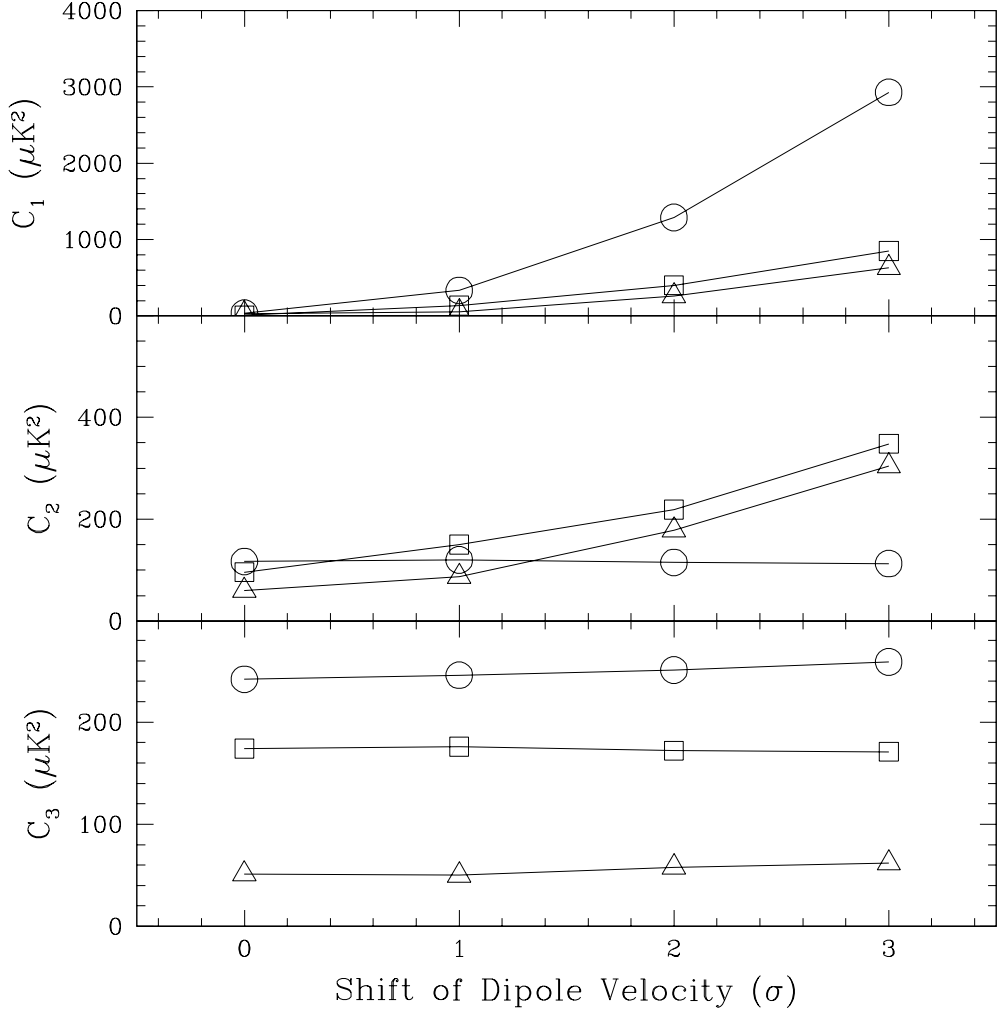}
\caption{
Uncorrected Kp2 cut-sky power (circles) and uncorrected
Kp2+hemisphere cut-sky power (triangles and boxes for north and
south respectively) for
the dipole (top), quadrupole (middle), and octopole (bottom), 
as a function of change in the magnitude of the dipole velocity 
(1$\sigma$~=~1.87 km s$^{-1}$).
The assumed NP is the galactic NP.  This figure shows that changing the
dipole velocity acts to increase the Kp2+hemisphere cut-sky power
for the quadrupole, so that asymmetry approaches unity and becomes
insignificant.  It also demonstrates that for
the Kp2 cut-sky, the effect of shifting the velocity is 
concentrated at the dipole, as expected.
}
\label{fig:uncor_quad}
\end{figure}

To determine the effect of using the Master algorithm upon quadrupole
power, we examine both uncorrected and corrected
cut-sky power spectrum estimates.
As illustrated in Figure \ref{fig:uncor_quad},
the uncorrected Kp2+hemisphere cut-sky power spectrum estimates for
the quadrupole (triangles and boxes for the north and south hemispheres
respectively in the middle panel)
are greatly affected when we increase the dipole velocity, with
the ratio of estimates tending toward unity (thus reducing the
significance of asymmetry).
The subsequent shifting of dipole power to the quadrupole
via mode coupling in the Master algorithm changes the quadrupole
power by $\lesssim$ 10\% (varying slightly as a function of direction).  
We conclude that the observed reduction in the significance of
asymmetry is primarily due to increased quadrupole power in each
hemisphere, and is not purely an artifact of the Master algorithm.

It is important to note that
increasing the sky coverage reduces the effect of dipole velocity
upon quadrupole power, and if we use the Kp2 mask alone, the effect is minimal 
(circles, middle panel).
Thus changing the dipole velocity has negligible impact on ``full sky"
power spectra, and does not help to explain any supposed irregularities
in these spectra (see, e.g., Land \& Magueijo 2005b and Vale 2005).
It is also important to note that
the increase in the dipole power for the Kp2 cut sky
at $\gtrsim$~1$\sigma$ (top panel, Figure \ref{fig:uncor_quad})
does not necessarily rule out higher velocities, as
the offsets of the tested
velocity from the true velocity and from the {\it COBE} DMR velocity
will have a non-negligible effect on the dipole power estimate, and
the contribution of the primordial dipole component is unknown but
could be significant ($\sim$ 1000 $\mu$K).

\begin{figure}[p]
\centering
\includegraphics[width=6.0in]{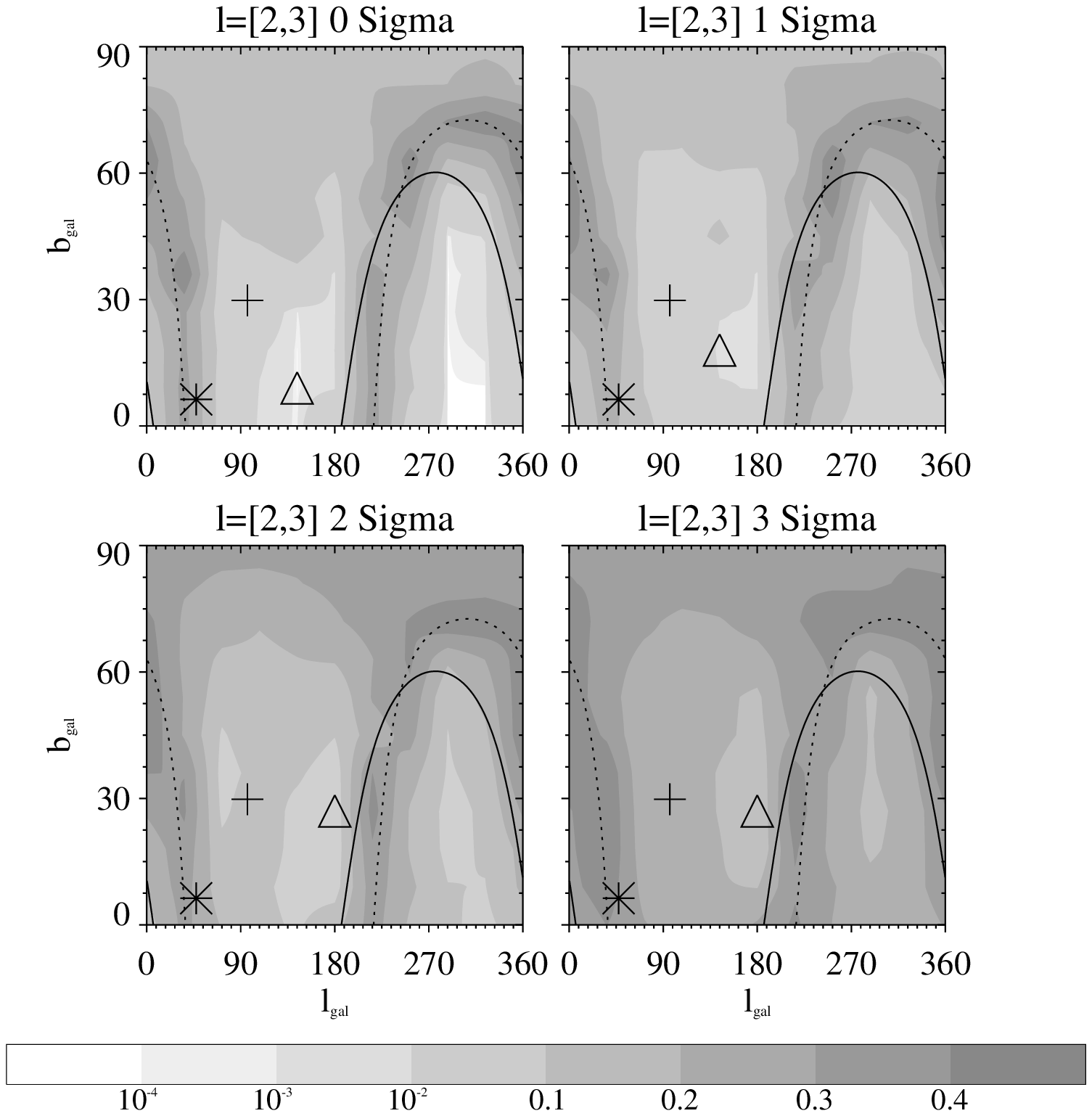}
\caption{
Same as the upper left panel of Figure \ref{fig:gc}, with
the magnitude of the dipole velocity increased by 1$\sigma$ (=~1.87 km s$^{-1}$),
2$\sigma$, and 3$\sigma$ (upper right, lower left, and lower right,
respectively).  The effects of increasing the dipole velocity are
to make $\alpha_{\rm global}$ insignificant, to make the direction of
maximum significance migrate toward the ecliptic plane, and to 
make the apparent great circle less visually striking.
}
\label{fig:contour_vel}
\end{figure}

In Figure \ref{fig:contour_vel}, we display how the contours of 
directional significance evolve as the magnitude of the dipole
velocity increases.  In addition to an increase in $\alpha_{\rm global}$,
we observe that the direction of maximum significance migrates 
toward the ecliptic plane, with the great circle still evident,
though not as strongly, at 3$\sigma$.

\paragraph{Loss Imbalance.}
We set $x_{\rm im}$ in eqs.~(\ref{eqn:ta}) and (\ref{eqn:tb}) to zero.
This has no discernible effect on asymmetry.

\paragraph{Masking.} 
To gauge the effect of masking, we make maps with no mask at all.
This has virtually no effect on asymmetry, presumably because the
scale size of the aforementioned galactic echoes is smaller than those
probed in our analysis.

\paragraph{Interpolation to Determine Spacecraft Orientation.}
The quaternions that encode spacecraft orientation are recorded at the
beginning of each 1.536 s science frame, while data are sampled
up to 30 times per frame.  Interpolation is normally done using Lagrange
interpolating polynomials (see the Appendix for details).
We test the effect of making maps with no interpolation
and with linear interpolation.
Turning off interpolation leads to pointing errors on the order of degrees,
and asymmetry becomes
more pronounced, with the significance increasing by two orders of 
magnitude ($\alpha_{\rm global} \sim$~10$^{-6}$ for $l$~=~[2,64]).
The use of linear interpolation has virtually no effect on asymmetry.
This indicates that the {\it WMAP} team's use of Lagrange interpolating
polynomials is a conservative algorithmic choice, and that
the small
errors in the recorded orientation of the spacecraft and the determination
of instrumental boresights ($\lesssim$ 1$\arcmin$; see
{\S}3.3.1 and 3.3.2 of Hinshaw I)
have little to no effect on asymmetry.

\paragraph{Planetary Cut.}
If a planet is determined to be 
within $\theta_{\rm cut}$~=~1.5$^{\circ}$ of either radiometer's normal vector, 
the temperature map is not updated for either.  
This value is conservative, being over four
times the FWHM of the V-band radiometer beam size (this helps mitigate
the effect of non-Gaussian shoulders in the beam profile).  We 
test the effects of making maps assuming $\theta_{\rm cut}$~=~0.5$^{\circ}$,
and $\theta_{\rm cut}$~=~0.0$^{\circ}$.  In both cases,
asymmetry becomes more pronounced over the full range $l$~=~[2,64].
If $\theta_{\rm cut}$~=~0.5$^{\circ}$, the effect upon
significances is $\lesssim$ 10\%, while if $\theta_{\rm cut}$~=~0.0$^{\circ}$,
$\alpha_{\rm global} \rightarrow$ 0.  The former result indicates
that reasonably changing $\theta_{\rm cut}$
would have virtually no effect on asymmetry.

\paragraph{Foreground Maps.}
We test the effect of foreground corrections by changing the relative
normalizations of the three foreground maps (dust, H$_{\alpha}$, and
synchrotron; see Bennett II for details).  We find that small shifts
($\sim$ 10\%) have little to no effect upon asymmetry,
which is not unexpected because of the markedly reduced level of
foreground emission outside the Kp2 mask with respect to that in the galactic 
plane.  (Because of the large number of normalization combinations tested,
we do not display our results in Table \ref{tab:map2to64ri}.)
To have a discernible effect, order-of-magnitude changes in the normalizations
are required.  This demonstrates that asymmetry is not simply related to 
uncertainties in relative normalization.

\section{Summary and Conclusions}

\label{sect:conc}

In this work, we analyze first-year {\it WMAP} data
to determine the significance of asymmetry in summed power between
opposite arbitrarily defined hemispheres.  We perform this analysis
on maps that we compute directly from the calibrated time-ordered
data, using a map-making algorithm that we have developed to be
similar to that used by the {\it WMAP} team.  By creating our
own maps, we are able to determine
whether altering elements of the map-making algorithm is sufficient
to affect the observed asymmetry (and perhaps to render it statistically
insignificant).

We follow Eriksen I and Hansen I by analyzing co-added V- and W-band foreground
corrected maps, and like these groups we find that there is a significant
difference in summed power between arbitrarily defined northern and
southern hemispheres.  We compute significance by simulating data from
the running-index LCDM power spectrum, determining the most significant
deviation from expectation in each dataset, fitting an extreme-value
distribution to these deviations, and computing the tail integral of
the best-fit distribution.
Over the multipole range $l$~=~[2,64],
we estimate the global significance of asymmetry to
be $\alpha_{\rm global} \sim$ 10$^{-4}$, and we find that this
value is not sensitive either to frequency (as determined by examining
co-added data in the Q, V, and W bands, and within each radiometer alone),
or to the details of the power spectrum used as input to simulations.  
These results
clearly indicate that the posited power asymmetry in {\it WMAP} data is real.
We determine the smallest 
multipole ranges for which $\alpha_{\rm global} \leq$~0.05; we find
twelve such ranges, including $l$~=~[2,3] and [6,7], 
for which $\alpha_{\rm global} \rightarrow$ 0.
Contour plots made for these ranges indicates an association between
the direction of maximum asymmetry and the ecliptic plane (four of
twelve directions within $\approx$ 4$^{\circ}$ and eight
within $\approx$ 20$^{\circ}$; $p_{\rm binomial} \lesssim$ 0.01),
although one must be mindful of the coarseness of our coordinate grid.
At large scales, the contours of least significance 
($\alpha_{\rm dir} \approx$
0.2-0.5) may be fit by eye with great circles inclined relative to the
ecliptic plane,
with inclinations of $\approx$ 30$^{\circ}$ and 90$^{\circ}$
for $l$~=~[2,3] and [6,7] respectively.  
We find that 
the $l$~=~[2,3] great circle is insensitive to frequency and
that it passes over preferred 
quadrupole and octopole axes derived by
de Oliveira-Costa et al.~and Schwarz et al.,
as well as the pointing direction of the CMB dipole.

We test the robustness of our results by using examining the
effect of map-making algorithmic assumptions upon asymmetry.
We create and analyze new maps after changing the monopole temperature, 
the magnitude of the dipole velocity,
the masks, the method of interpolation
used to determine the pointing direction of radiometer normal
vectors as a function of time, the planetary cut, and
the relative normalizations of the foreground maps. 
In virtually all cases, the alterations had little to no effect
upon asymmetry either for the range $l$~=~[2,64] or at the largest
scales.  An
exception to this is the magnitude of the dipole velocity: we find that 
increasing it has a marked effect on the direct (i.e., uncorrected) estimate of 
quadrupole power in the Kp2+hemisphere cut skies, increasing power
in both the northern and southern hemispheres such that asymmetry
approaches unity (and becomes insignificant).  
Changing the magnitude of the dipole velocity 
is a frequency-independent change, consistent with observations
that the significance of asymmetry is 
frequency-independent.
If the true dipole
velocity magnitude is $\approx$ 1-3$\sigma$ larger than the current value,
large-scale asymmetry may disappear.  However, it will not influence asymmetry
estimates at smaller scales, nor appreciably change
full-sky power spectra.
This is consistent with the conclusion of Hansen I 
that whatever causes asymmetry at large scales is not what causes it
at smaller scales; however, they posit that 
galactic contamination is the cause of large-scale asymmetry.

The effect of changing the magnitude of the dipole velocity, along
with the observation of structures in our contour plots that follow
great circles inclined relative to the ecliptic plane,
suggests that the observed power asymmetry
may be (at least partially) caused by the use of
an incorrect dipole vector in combination with
a systematic or foreground process that is associated with the ecliptic.
The natural next steps are to determine the dipole velocity vector that 
minimizes asymmetry (with error bars), to make new maps based on that vector, 
and to examine the frequency dependence of any remaining significant asymmetry
to attempt to differentiate between systematic or foreground causes.
However, without the ability to 
recalibrate the first-year {\it WMAP} data, such an exercise
may not yield accurate results.  We thus would ask that both raw 
time-ordered data
and calibration software be made available to the public in future
data releases.

\acknowledgements
The authors would like to thank Gary Hinshaw and Paul Butterworth
of the {\it WMAP} team for their great help in understanding the {\it WMAP}
map-making algorithm and for providing to us the {\it WMAP} quaternion
interpolation software.  
We would also like to thank Tony Banday,
Hans Kristian Eriksen, Frode Hansen, and Kris G\'orski for reviewing
an earlier version of this text and providing very helpful comments.  
RCN thanks Rob Crittenden for useful discussions.
Our parallelized map-making software was run on the NCSA/Teragrid Linux
Cluster (grant number MCA04N009, PI Roy Williams), 
supported by the National Science Foundation 
under the following NSF programs: Partnerships for 
Advanced Computational Infrastructure, Distributed Terascale Facility (DTF) 
and Terascale Extensions: Enhancements to the Extensible Terascale Facility.
This work was also supported by NSF grants ACI-0121671, Statistical
Data Mining in Cosmology, and AST-0434343, Nonparametrical
Statistical Methods for Astrophysical and Cosmological Data.

\appendix

\section{Map-Making Recipe}

\label{sect:mapapp}

In this Appendix, we lay out the details of our map-making algorithm,
summarized in {\S}\ref{sect:map}.

\subsection{Preliminaries}

We begin by assuming two values:
the velocity of the Sun relative to the CMB rest frame 
in galactic coordinates,
$(v_{\odot,\rm gal}^x,v_{\odot,\rm gal}^y,v_{\odot,\rm gal}^z)$~=~($-$26.26,$-$243.71,274.63) km s$^{-1}$ (Bennett I), 
and the monopole temperature $T_{\rm o}$~=~2.725 K (Mather et al.).
We input the planetary cut angle $\theta_{\rm cut}$ (default 1.5$^{\circ}$), 
the method of quaternion interpolation (none, linear, or using Lagrange
interpolating polynomials, with the latter the default), 
and the number of iterations (default 20).

We input the mask from {\tt wmap\_composite\_mask\_yr1\_v1.fits}.  The
masking column in this file is bit-coded; we examine bit 5 to set
the processing foreground mask, which is the Kp8 mask with no edge
smoothing.  We do not apply the point source mask, as we expect point
sources to have little effect on results for $l \leq$~64.

We then read in a list of galactic coordinates for each {\tt HEALPix} pixel,
generated once using the IDL routine {\tt healpix\_nested\_vectors}.
We assume {\tt nside}~=~512 (or 3,145,728 map pixels).  

\subsection{Making Map Estimates From Time-Ordered Data}

The making of maps involves four embedded loops: looping over iteration
number, looping over the 366 time-ordered data (TOD) files,\footnote{
\scriptsize {\tt http://lambda.gsfc.nasa.gov/product/map/dr1/map\_tod.cfm}}
looping over
the science frames within each TOD file, and looping over the data
vector contained within each science frame.

During each loop, we open and read each TOD file, as opposed to keeping
their contents in memory after a first reading.  
(This is not strictly necessary; see {\S}\ref{sect:comp} below.)
Each TOD file contains five tables, of which we are
interested in three: the Meta Data, Line-of-Sight (LOS), and Science Data
tables.

Each Meta Data table contains 1875 rows (recorded every 46.08 s, or
30 science frames).
We extract from this table elements of the time column, the spacecraft 
position column (needed for determining planetary positions), the
spacecraft velocity column, and the column containing the
33 $\times$ 4 quaternion matrices that encode the spacecraft orientation for
each of science frame (the extra 3 elements provide seamless
interpolation; see below).  The spacecraft position and velocity are
given in celestial coordinates; they are converted to galactic coordinates
as detailed below.

The LOS table contains the orientation of each radiometer horn in
spacecraft coordinates, which we extract once.  This information, combined
with the information from the quaternion matrices, allows us to determine
the normal vector for each radiometer horn in galactic coordinates,
and thus the map pixel numbers associated with each radiometer as a 
function of time.  We return to this below.

Each Science Data table contains 56250 rows (30$\times$1875), corresponding
to science frames of length 1.536 s.  From each row, we extract a 
data vector of length $N_{\rm data}$ 
(=15 for Q-band data, 20 for V-band data, and 30 for W-band data), 
and a status code.
If this code is odd (i.e., if bit 1 is set), the data within the
vector are considered bad and are not used.

For each datum $d$, an interpolated quaternion is computed, 
using a Lagrange interpolating polynomial formulation
provided by the {\it WMAP} team (G.~Hinshaw \& P.~Butterworth, private
communication).  This formulation requires
the quaternions for the previous, current, and following two science frames.
We calculate the time ${\Delta}t$ between that of the datum and the
beginning of the current science frame, and convert this to a
fractional offset $f~=~{\Delta}t{\slash}$1.536 s.
We then define four weights:
\begin{eqnarray}
w_1~&=&~-f(f-1)(f-2)/6 \nonumber \\
w_2~&=&~(f+1)(f-1)(f-2)/2 \nonumber \\
w_3~&=&~-(f+1)f(f-2)/2 \nonumber \\
w_4~&=&~(f+1)f(f-1)/6 \nonumber \,,
\end{eqnarray}
and determine the four elements of the interpolated quaternion:
\begin{equation}
q_{\rm int,\it i}~=~\sum_{j=1}^{4} w_j q_{j,i} \,,
\end{equation}
where $j$ represents a science frame.
The next step is to normalize $q_{\rm int}$ and to convert it 
to a transformation matrix
(here we drop the ``int" subscript):
\begin{equation}
{\bf T}~=~\left( \begin{array}{ccc} q_1^2-q_2^2-q_3^2+q_4^2 & 2(q_1q_2-q_3q_4) & 2(q_1q_3+q_2q_4) \\ 2(q_1q_2+q_3q_4) & -q_1^2+q_2^2-q_3^2+q_4^2 & 2(q_2q_3-q_1q_4) \\ 
 2(q_1q_3-q_2q_4) & 2(q_2q_3+q_1q_4) & -q_1^2-q_2^2+q_3^2+q_4^2 \end{array} \right)
\end{equation}
This transformation matrix is in turn used to convert the radiometer horn 
normal vector $n_{\rm sp}$ from spacecraft to celestial coordinates:
\begin{eqnarray}
{\bf N_{\rm cel}}~&=&~{\bf T} \left( \begin{array}{ccc} n_{1,\rm sp} & n_{2,\rm sp} & n_{3,\rm sp} \\ n_{1,\rm sp} & n_{2,\rm sp} & n_{3,\rm sp} \\ n_{1,\rm sp} & n_{2,\rm sp} & n_{3,\rm sp} \end{array} \right) \\
n_{i,\rm cel}~&=&~\sum_{j=1}^3 N_{i,j,\rm cel} \,.
\end{eqnarray}
We then convert to galactic coordinates using the transformation matrix
\begin{equation}
{\bf T_{\rm cg}}~=~\left( \begin{array}{ccc} \cos{\Psi}\cos{\Phi}-\sin{\Psi}\sin{\Phi}\cos{\Theta} & \sin{\Psi}\cos{\Phi}+\cos{\Psi}\sin{\Phi}\cos{\Theta} & \sin{\Phi}\sin{\Theta} \\ -\cos{\Psi}\sin{\Phi}-\sin{\Psi}\cos{\Phi}\cos{\Theta} & -\sin{\Psi}\sin{\Phi}+\cos{\Psi}\cos{\Phi}\cos{\Theta} & \cos{\Phi}\sin{\Theta} \\ \sin{\Psi}\sin{\Theta} & -\cos{\Psi}\sin{\Theta} & \cos{\Theta} \end{array} \right) \,,
\end{equation}
where $\Psi$~=~282.85948$^{\circ}$, $\Phi$~=~327.06808$^{\circ}$, and
$\Theta$~=~62.871750$^{\circ}$.  The transformation is:
\begin{equation}
{\vec n}_{\rm gal}~=~{\bf T_{\rm cg}} {\vec n}_{\rm cel} \,.
\end{equation}
The mathematics of the subsequent transformation from ${\vec n}_{\rm gal}$
to {\tt HEALPix} pixel number is complex and will not be reproduced here;
our software routine is based upon the {\tt HEALPix} IDL routine
{\tt vec2pix\_nest}.

At this point, we know to which pixels the $A$ and $B$ radiometer horns are
pointing.  The next step is to determine
the Doppler shift of the monopole relative to the spacecraft.
The spacecraft velocity is recorded every 46.08 seconds, as noted above.
Because the velocity varies little during that amount of time, 
it is not strictly necessary to interpolate to determine velocities
for each datum.  (However, we have made the choice
to linearly interpolate the velocities.)  We convert the velocity 
associated with the datum from celestial to galactic coordinates:
\begin{equation}
{\vec v}_{\rm sp,gal}~=~{\bf T_{\rm cg}} {\vec v}_{\rm sp,cel} \,.
\end{equation}
Now we can remove the contribution of the Doppler-shifted monopole from
the datum (here, we drop the ``gal" subscript):\footnote{
What we denote ${\Delta}T_{{\rm CMB},t}$ actually consists of contributions
from the CMB and foregrounds.  One can correct for the
foregrounds during mapping
by subtracting
$T_{\rm A}^{\rm FG}-T_{\rm B}^{\rm FG}$ 
and (if desired)
$T_{\rm A}^{\rm FG}({\vec \beta}_{\rm sp} \cdot n_{\rm A})-T_{\rm B}^{\rm FG}({\vec \beta}_{\rm sp} \cdot n_{\rm B})$
from ${\Delta}T_{{\rm cal},t}$, along with the contribution of the 
Doppler-shifted monopole.  The latter factor assumes that the 
foreground emitters are stationary in the barycenter frame of reference.
We find that the difference between performing foreground 
corrections during mapping
and after mapping is negligible ($\lesssim$ 1 $\mu$K in each sky map pixel).
}
\begin{eqnarray}
{\vec \beta}~&=&~\frac{1}{c}({\vec v}_{\rm sp} + {\vec v}_{\odot}) \\
{\Delta}T_{{\rm CMB},t}~&=&~{\Delta}T_{{\rm cal},t}-T_{\rm o} \times \left[{\vec \beta} \cdot ({\vec n}_{\rm A}-{\vec n}_{\rm B}) + ({\vec \beta} \cdot {\vec n}_{\rm A})^2 - ({\vec \beta} \cdot {\vec n}_{\rm B})^2 \right] \,,
\end{eqnarray}
and update the temperature map and the counter recording how many times
each pixel is observed:
\begin{eqnarray}
n_{{\rm obs},p_A} &=& n_{{\rm obs},p_A}+w_t \\
{\Delta}T_{{\rm CMB},p_A,i+1} &=& {\Delta}T_{{\rm CMB},p_A,i+1} + w_t\frac{{\Delta}T_{{\rm CMB},t}+(1-x_{\rm im}){\Delta}T_{{\rm CMB},p_B,i}}{1+x_{\rm im}} \,,
\end{eqnarray}
and
\begin{eqnarray}
n_{{\rm obs},p_B} &=& n_{{\rm obs},p_B}+w_t \\
{\Delta}T_{{\rm CMB},p_B,i+1} &=& {\Delta}T_{{\rm CMB},p_B,i+1} + w_t\frac{-{\Delta}T_{{\rm CMB},t}+(1+x_{\rm im}){\Delta}T_{{\rm CMB},p_A,i}}{1-x_{\rm im}} \,,
\end{eqnarray}
where $i+1$ is the current iteration (${\Delta}T_{p,0}$ is the input
zero-temperature map).
The counter is actually updated during only the first iteration as
$n_{{\rm obs},p}$ does not change from iteration to iteration.
As noted in {\S}\ref{sect:map}, 
we assume the statistical weight $w_t$~=~1.0 for all data, as the
first-year data release does not contain information on the statistical
weighting model used by the {\it WMAP} team.  At the end of a given
iteration, we divide by $n_{{\rm obs},p}$ to determine the weighted
average.

Equation (\ref{eqn:ta}) is skipped (i.e., the map is not updated) if pixel
$B$ lies within the processing mask, while equation (\ref{eqn:tb}) is skipped
if pixel $A$ is within the mask.  Both equations are skipped if either
pixel $A$ or $B$ is within an angle of $\theta_{\rm cut}$ degrees of any outer planet
(except Pluto; see {\S}\ref{sect:planet} below).

\begin{figure}[p]
\centering
\vskip -0.5in
\includegraphics[width=4.0in]{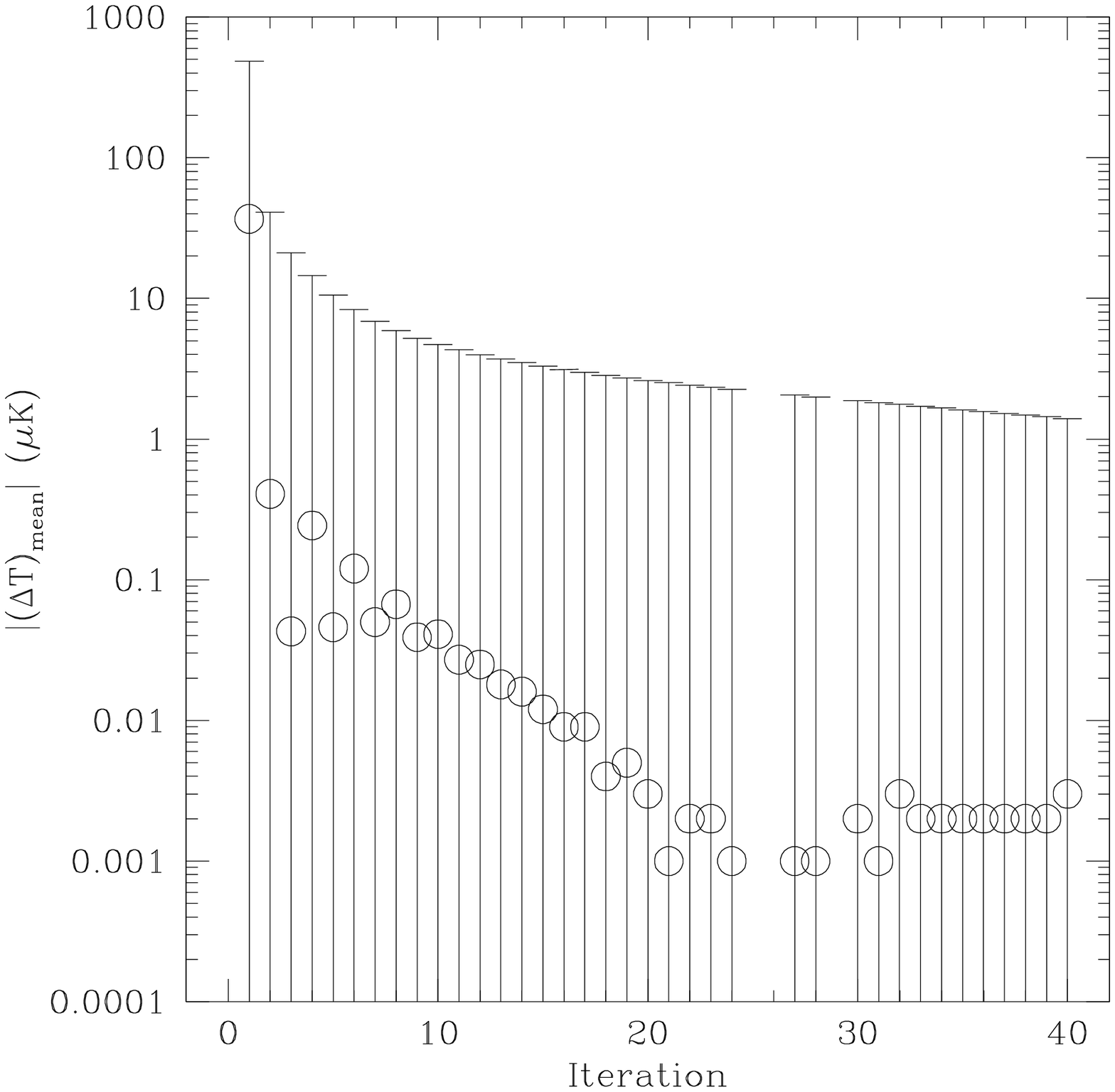}
\hfill
\vskip -0.25in
\includegraphics[width=4.0in]{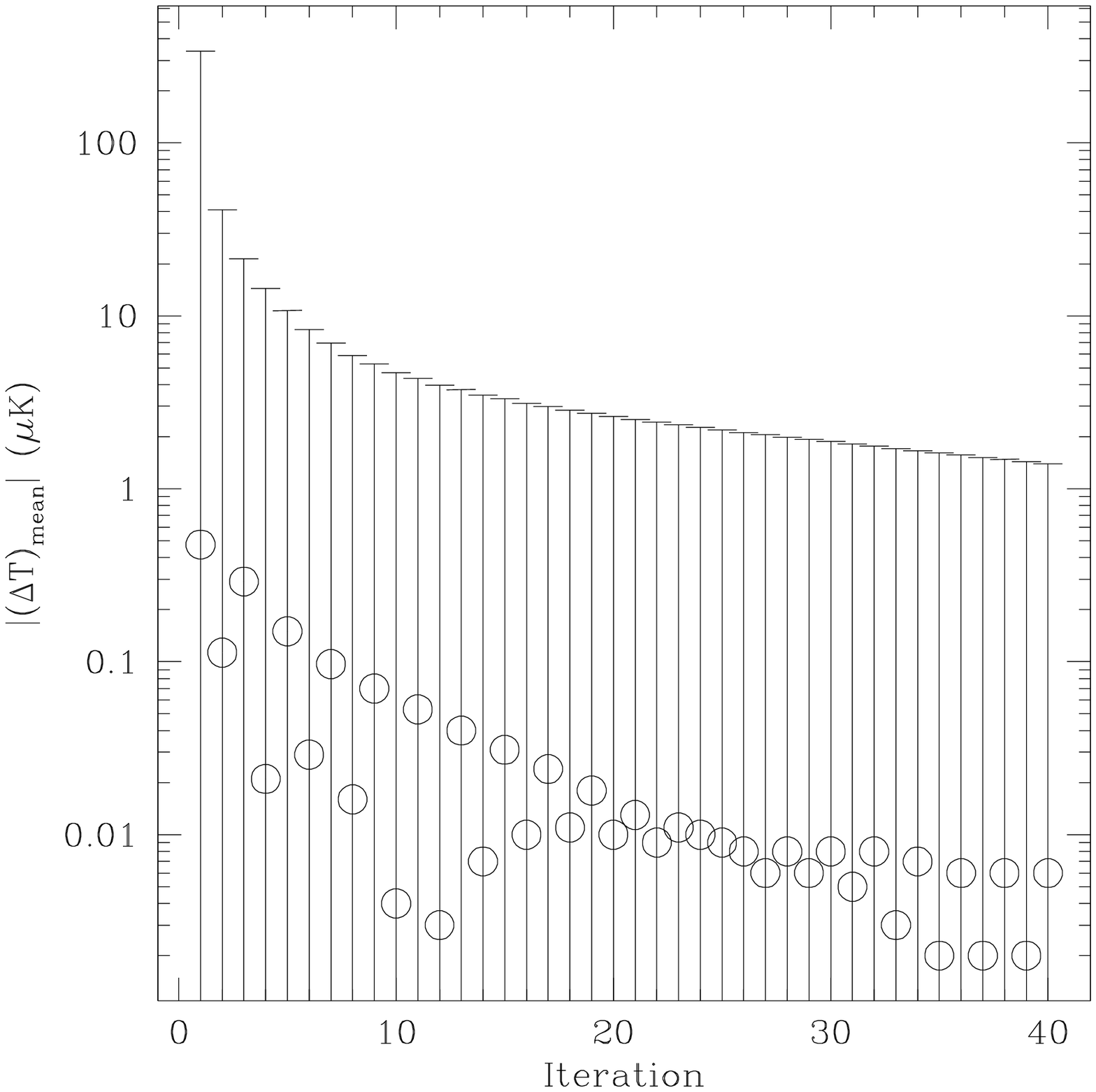}
\vskip -0.25in
\caption{The average change in temperature in each map pixel from one
iteration to the next, as a function of iteration, for all pixels
(top) and for pixels outside the Kp8-without-edge-smoothing
mask (bottom).  The particular values shown here
were generated for a 40-iteration W1 radiometer map.  The 1$\sigma$ error
bars indicate the size of the distributions of ${\Delta}T$, and are
not estimates of the errors on the mean.  These plots
demonstrate the robustness of qualitative conclusions based on 
20-iteration maps.}
\label{fig:conv}
\end{figure}

Following Hinshaw I, we generally iterate 
20 times; in Figure \ref{fig:conv}, we show that this number of iterations is
sufficient for convergence.

The map that is generated after 20 iterations is not an absolute map,
but a difference map with no precisely defined zero-point.
The {\it WMAP} team, which like us uses a zero-temperature input map,
determines the zero-point by fitting a function of the form
$T~=~C_1{\csc}b_{\rm gal}+C_0$ to southern 
galactic hemisphere data below $b_{\rm gal}$~=~$-$15$^{\circ}$
for each map (Bennett I).  
We apply a constant offset to
our maps such that the mean pixel temperature
matches that of the {\it WMAP} maps.  

\subsection{Planetary Cuts}

\label{sect:planet}

In theory, we should determine the positions of all outer
planets in each science frame.  However, position determination
is a major computational bottleneck.  Hence whenever we
determine planetary positions, we
estimate the length of time necessary for {\it WMAP} to rotate through
the smallest angle between either radiometer normal vector and any
outer planet, and
we do not determine positions for that length of time.
This scheme is unduly
conservative, since we do not check to see if {\it WMAP} is rotating
{\it toward} that planet, but it effectively eliminates the bottleneck. 

We compute planetary positions using the publicly
available {\tt axBary} software package\footnote{\scriptsize
{\tt http://heasarc.gsfc.nasa.gov/listserv/heafits/msg00050.html.}}.
We use the package function {\tt dpleph}, based on the ephemeris
{\tt JPLEPH.405}.  For each outer planet, we (a) determine its position
relative to the spacecraft in celestial coordinates for Julian
date $t_{JD}$, applying
spacecraft positions read in from the Meta Data table; (b) update
that position given the actual time of the science frame; (c) convert
that position to galactic coordinates using the transformation
matrix ${\bf T_{\rm cg}}$; and 
(d) calculate $\theta_{\rm A/B}={\rm acos}({\vec n}_{\rm planet}\cdot{\vec n}_{\rm A/B,gal})$.
Note that we do not include the effect of light travel time in our algorithm,
as the planets move by less than one sky pixel ($\approx$ 7$\arcmin$)
during that period.
If $\theta_{\rm A}$ or $\theta_{\rm B}$ is $\leq \theta_{\rm cut}$ 
(default 1.5$^{\circ}$), we
do not update the temperature map for either pixel.

\subsection{Computational Details}

\label{sect:comp}

Map-making is a computationally intensive task.  
Given current typical CPU speeds ($\sim$ GHz),
making a single 20-iteration map takes $\sim$ 1 CPU day.
To reduce map-making time, we parallelized our C code
using the Message Passing Interface (MPI) library.\footnote{\scriptsize
{\tt http://www-unix.mcs.anl.gov/mpi}
}  In this instance, parallelization is particularly easy:
each processor is assigned to loop over a subset of TOD files,
taking the estimated map from the previous iteration and 
updating it only for that subset.  At the end of each iteration,
the updated maps are broadcast to the root node, which sums
them; the summed map is then broadcast back to the processing
nodes for use during the next iteration.
We run our parallelized code on the Teragrid Linux Cluster 
({\tt www.teragrid.org}), which links massively parallel supercomputing
clusters at nine sites, including NCSA, where we do the bulk
of our map-making and where a typical 24-processor, 20-iteration run takes
$\approx$ 25 CPU minutes.  We note that although we read in the 
data sequentially during each iteration, because each processing
node has its own memory we could in theory keep the time-ordered data in 
memory after being read once.  The total amount of, e.g., W-band data
that would need to be kept in memory would be reduced from 
366 days $\times$ 56,250 data rows $\times$ 30 data =
6.2 $\times$ 10$^8$ ($\approx$ 2.5 Gb of memory) to, say,
2.6 $\times$ 10$^7$ ($\approx$ 100 Mb of memory)
for 24 processors, an amount that each node (containing two processors)
could easily handle.

\vfill\eject

\end{document}